\documentclass[useAMS,usenatbib]{mn2e}
\usepackage{graphicx}
\usepackage{epsfig}
\usepackage{natbib}  
\usepackage{epstopdf}
\newcommand\aj{{AJ}}%
\newcommand\apj{{ApJ}}%
\newcommand\apjl{{ApJ}}%
\newcommand\apjs{{ApJS}}%
\newcommand\aap{{A\&A}}%
\newcommand\mnras{{MNRAS}}%
\newcommand\nat{{Nature}}%
\def\simgt{\lower.5ex\hbox{$\; \buildrel > \over \sim \;$}}
\def\simlt{\lower.5ex\hbox{$\; \buildrel < \over \sim \;$}}

\def\c12{$^{12}$C}
\def\neo20{$^{20}$Ne}
\def\al27{$^{27}$Al}
\def\ne22{$^{22}$Ne}
\def\na23{$^{23}$Na}
\def\mg25{$^{25}$Mg}
\def\mag26{$^{26}$Mg}

\def\tend{$t_{\rm end}$}
\def\taupeak{$t_{\rm acc}$}
\def\sigpri{$\tau_{\rm pri}$}

\newcommand{\msun}{\ensuremath{\, {M}_\odot}}
\newcommand{\Msun}{\ensuremath{\, {M}_\odot}}
\newcommand{\ocen}{$\omega$~Cen}

\newcommand{\pristine}{$_{\rm pristine}$}
\title[Models for the lithium abundances of multiple populations in globular clusters]{Models for the lithium abundances of multiple populations in globular clusters and the possible role of the Big Bang lithium
}
\author[F. D'Antona, A. D'Ercole, R. Carini,  E. Vesperini, \& P. Ventura]
{Francesca D'Antona$^{1}$, Annibale D'Ercole$^{2}$, Roberta Carini$^{1,3}$, Enrico Vesperini$^{4}$,\newauthor \& Paolo Ventura$^{2}$
\thanks{E-mail: dantona@oa-roma.inaf.it (FD)}
\\
$^{1}$ INAF, Osservatorio Astronomico di Roma, Via Frascati 33, I-00040 Monteporzio Catone (Roma), Italy.\\
$^{2}$  INAF- Osservatorio Astronomico di Bologna, via Ranzani 1, I-40127 BOLOGNA (Italy).\\
$^{3}$ Department of Physics, Universit\`a di Roma ``La Sapienza'', Roma, Italy\\
$^{4}$ Department of Physics, Drexel University, Philadelphia, PA 19104, USA
}
\begin{document}

\date{Accepted . Received ; in original form }

\pagerange{\pageref{firstpage}--\pageref{lastpage}} \pubyear{2011}

\maketitle

\label{firstpage}

\begin{abstract}

Globular cluster stars show chemical abundance patterns typical of hot--CNO processing, and photometric evidence for the presence of multiple populations. Lithium  is easily destroyed by proton capture in stellar environments, so determining its abundance may be crucial to discriminate among different models proposed to account for the origin of the gas from which the multiple populations form. 
In order to reproduce the observed O--Na anticorrelation and other patterns typical of multiple populations, the formation of second generation stars must occur from the nuclearly processed stellar ejecta, responsible of the chemical anomalies, {\it diluted} with pristine gas having the composition of first generation stars. This gas is either a remnant of the first phases of star formation, or it has been re-accreted on the cluster core after the end of the SN II explosions. As a consequence, the lithium abundance in the unprocessed gas ---which is very likely to be equal to the lithium abundance emerging from the Big Bang--- affects the lithium chemical patterns among the cluster stars.
This paper focuses on a scenario in which processed gas is provided by asymptotic giant branch (AGB) stars. We examine the predictions of this scenario for the lithium abundances of multiple populations.
We study the role of the non--negligible lithium abundance in the ejecta of massive AGB (A(Li)$\sim$2), and, at the same time, we explore how our models can constrain the extremely large ---and very model dependent--- lithium yields predicted by recent super--AGB models. We show that the super--AGB yields may be tested by examining the lithium abundances in a large set of blue main sequence stars in \ocen\ and/or NGC~2808. 
In addition, we examine the different model results obtained by assuming for the pristine gas either the Big Bang abundance predicted by the standard models (A(Li)=2.6--2.7), or the abundance
detected at the surface of population II stars (A(Li)=2.2--2.3), and we  show that, once a chemical model is well constrained by a comparison with the observations, the O--Li distribution could perhaps be used to shed light on the primordial lithium abundance. 
\end{abstract}


\section{Introduction}
\label{sec:intro}

The spread of light elements abundances in globular clusters is larger
than in field stars of similar metallicity and requires the presence
of a population of stars formed from matter processed through the hot CNO cycle and by other proton--capture reactions on light nuclei (hereafter we will refer to this population as second generation, SG).  The
site of processing of this matter in the first generation (FG) stars of the proto--cluster is a subject of intensive investigation. The two
main scenarios are:  the ``AGB  scenario", in which the site of
processing are the hot bottom burning envelopes of massive asymptotic
giant branch (AGB) stars \citep{ventura2001, dc2004, karakas2006}, and
the  ``fast rotating massive stars (FRMS) scenario"
\citep{prantzos2006,   meynet2006, decressin2007a}, in which the site
is the interior of fast rotating massive stars. 

Problems are present in both scenarios \citep[see, e.g.,][for a critical review]{renzini2008}. In all the scenarios proposed so far, the composition of SG stars is, in most cases,  best explained by dilution of the polluting ejecta with pristine gas \citep[see, e.g.][]{dercole2011}. 
One possible discriminant between the two scenarios is the comparison between the abundance of lithium in FG and SG stars. In fact,  massive stars destroy lithium, while  massive AGB produce it at the beginning of the HBB \citep{vdm2002}, so possibly the prediction of the two models differ. Also for other scenarios lithium can be a powerful discriminant: the polluting matter is Li--free also if it comes from runway collision between massive stars \citep{sills2010}, or non--conservative evolution of massive binaries \citep{demink2009},  while {\it the diluting gas may be Li--free,} if it comes from mass loss from FG stars by winds  \citep{gratton2010}, and also if it is made up by the matter in non conservative evolution of interacting close binaries \citep{vanbeveren2012} or stripped during close encounters in the cluster core \citep[][]{carini2012}.
 
Self--consistent models to study the chemical evolution of lithium in multiple-population  GCs  are still lacking, while some sets of observational data are already available. \cite{decressin2007b} show that the anticorrelation Li--Na in the data of  NGC~6752 \citep{pasquini2005} are fully compatible with a simple dilution model, but \cite{shen2010} present new data Li--O, probably  not compatible with dilution with Li--free matter. In other clusters, like NGC~6397 \citep{lind2009}, M~4 \citep{dorazi2010, monaco2012}, 47~Tuc \citep{dorazi47tuc2010} the FG and SG stars have very similar lithium content, and Li--depleted stars may be attributed to convective dilution. Naively, the similar abundances in these clusters seems in better agreement with the AGB scenario. 

In this paper, we study the lithium abundance patterns resulting from the chemical evolution models presented by \cite{dercole2010} (Paper I) and  \cite{dercole2012cev1} (Paper II). 
The models are based on the AGB scenario, and on the hydrodynamical computations by \cite{dercole2008}, and can reproduce the chemical patterns both in clusters having an extended O--Na anticorrelation, like NGC~2808, and in clusters showing a mild anticorrelation, like M~4. 

The outline of the paper is the following.
In Sect.~\ref{inputs} we briefly summarize the model, describe the lithium yields of the AGB and super--AGB models. As there is a still debated discrepancy between the predictions for primordial lithium from the standard Big Bang nucleosynthesis and the abundance observed in the atmospheres of population II stars, an additional hypothesis must be made, concerning the choice of the initial lithium abundance in the diluting pristine gas. 

Sect.~\ref{m4tot} compares the data by \cite{monaco2012}  for M~4 with the models that  account for the O--Na anticorrelation. 

In Sect.~\ref{2808} we show that the predictions for lithium in the
model  for NGC~2808 is relevant also for a comparison with the data by \cite{shen2010}  for NGC~6752. We will show that in this case the lithium yield of AGB stars is necessary to reproduce the observations, and a better determination of the O--Li patterns may provide an independent constraint on the lithium abundance emerging from  Big Bang. 

Sect.~\ref{li0} shortly examines the case of dilution of nuclearly processed ejecta with lithium--free gas.

The possible constraints posed by multiple population abundance patterns on the Big-Bang lithium abundance are further discussed in Sect.~\ref{bbli}. In Sect.~\ref{final} we summarize our conclusions.

\section{Inputs of the model}
\label{inputs}
\subsection{The chemical evolution model}
The chemical evolution model, described in detail in Paper I and Paper~II, assumes that the proto--GC becomes completely devoid of pristine gas at the end of the SN~II epoch, and begins a new star formation epoch when the low velocity winds of the super--AGB stars first, and of the massive AGB stars later on, collect in a cooling flow in the cluster core. Formation may occur directly from the super--AGB and AGB ejecta, but a phase of re-accretion of pristine gas (see e.g. D'Ercole et al. 2008, 2011)
may lead to dilution of the ejecta with the original matter. This phase is characterized by an epoch of maximum accretion of pristine gas, \taupeak, and a timescale of accretion \sigpri, that in our formulation is the standard deviation of the gaussian describing the process. A third parameter, the end (\tend) of the second star formation epoch, provides a further independent timing of the process. It is also possible to accumulate and mix the ejecta for a time $t_{\rm f}$\ before starting star formation. Further details on the general behavior of models and on the adopted symbolism can be found in Paper I.
In Paper II we introduced in the model the new yields for super--AGB models from 6.5 to 8\msun\ computed by \cite{vd2011}, that complement the results by \cite{vd2009}, to revise the results presented in Paper I. 
In this work we explore the multiple generation lithium distribution resulting from some of the models computed in Paper~II for the clusters M~4 and NGC~2808. The main parameters of the chosen models are summarized in Table~\ref{t1}. We then extend the model discussion to other clusters for which data of lithium abundances are available (NGC~6397, NGC~6752, \ocen).
\begin{table}
\caption{Dynamical input parameters for the simulations}             
\label{t1}      
\begin{tabular}{l c c c c c c c c cccccc }     
\hline       
Model & \taupeak  & \sigpri & \tend & $ t_{\rm f}$  &$\rho_{\rm *,FG}$  &$\rho_{\rm 0,pr}$ & $\nu$ & $x$ \\
\\
\hline                                                                            
M~4-0      &  65  &  2   &   105  & 0 & 9.4 &0.091 & 0.1 & 0.7  \\ 
M~4-2    &  43  &  2  &   60  & 0 & 940 &0.05 & 0.1 & 0.5 \\
M~4-3    &  43  &  2  &   50  &  45 &940 &0.04  &0.5 & 0.5 \\
M~4-4     &  43  &  2  &   55  & 45 &940 &0.04 & 0.5 & 0.5  \\
M~4-5    &  48  &  10  &   58  & 48 & 940 &0.08 & 0.5 & 0.5 \\
2808-1   &   65  &  8   &   90  & 0 & 240 &0.0095&1  & 0.4 \\
\hline                                                                                                       
\end{tabular}
\leftline{\taupeak : time at which maximum accretion of pristine matter occurs (Myr)}
\leftline{\sigpri : timescale of the pristine gas accretion process (Myr)}
\leftline{\tend: time at which the SG star formation ends (Myr)}
\leftline{$t_{\rm f}$ : time until which the AGB ejecta are accumulated before star}
\leftline{   formation begins} 
\leftline{$\rho_{\rm{*,FG}}$: densiy of FG stars (\msun pc$^{-3}$)}
\leftline{$\rho_{\rm{0,pr}}$: densiy  regulating the amount of available pristine gas,}
\leftline{    normalized to $\rho_{\rm *,FG}$}
\leftline{$\nu$:   star formation efficiency}
\leftline{$x$ = $\rho_{{\rm *,SG}}/\rho_{{\rm *tot}}$: ratio between the SG stars and the total}
\leftline{   nowadays alive stars ($\rho_{{\rm *,tot}} = \rho_{{\rm *,FG}} + \rho_{{\rm *,SG}}$)}
\end{table}

 \begin{figure}
\resizebox{1.\hsize}{!}{\includegraphics{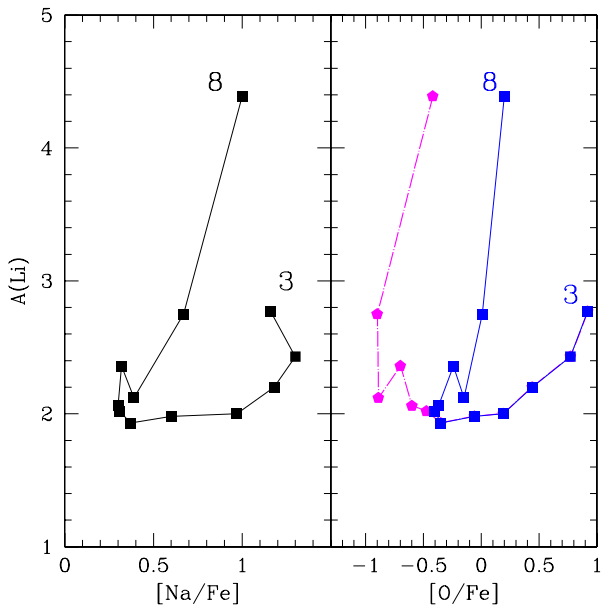}}
%
\caption{The full lines represents the lithium versus sodium (left), and lithium versus oxygen (right panel) abundances in the ejecta of super--AGB and AGB stars. Each dot represents an initial mass, the maximum and minimum mass are labelled. The dash--dotted line represent the oxygen abundance that would be seen in giants having very large helium abundance (for stars directly formed from the super--AGB ejecta), thanks to the 'in situ' deep mixing described in D'Ercole et al. (2012). Notice that we plot again the {\it initial} lithium abundances in the gas, but actually in these giants lithium is drastically reduced, both by standard convective mixing, and, even more, by the same convective deep mixing to which the oxygen reduction is attributed.
}
\label{f1} 
\end{figure}

\begin{table}
\caption{Lithium average abundances in the ejecta for Z=10$^{-3}$}             
\label{mloss}      
\centering          
\begin{tabular}{c c c c c}     
\hline\hline       
$M/M_{\odot}$ &  $\tau$/10$^6$yr & A(Li) & M$_{core}$/M$_\odot$ & core  \\ 
\hline         
3.00  &  332    & 2.77  & 0.76 & CO \\
3.50  &  229    &2.43  & 0.80 & CO  \\
4.00  &  169.5 &2.20  & 0.83 & CO \\
4.50  &  130.3 &2.00  & 0.86 & CO  \\
5.00  &  103.8 &1.98  & 0.89 & CO  \\
5.50  &  85.1   &1.93  & 0.94 & CO  \\
6.00  &  71.2   &2.02  & 1.00 & CO \\
6.30  &  65.2   &2.06  & 1.03 & CO \\
6.50  &  61.5   &2.36  & 1.09 & O-Ne\\
7.00  &  53.7   &2.12  & 1.20 & O-Ne\\
7.50  &  46.8   &2.75  &1.27 &  O-Ne\\
8.00  &  38.8   &4.39  & 1.36 &  O-Ne\\
\hline\hline
\end{tabular}
\label{t2}
\end{table}

\subsection{Lithium abundances in AGB and super--AGB ejecta}
For the sake of completeness, we list in Table~\ref{t2} the lithium abundances in the ejecta of AGB and super--AGB stars adopted in this work, and the evolutionary times of the corresponding masses. Notice that the assumptions on the initial lithium abundance in the gas forming the FG stars have no consequences on the yields. In fact, the initial lithium is fully depleted in the stellar envelopes, by the time the lithium production due to the Cameron-Fowler (1971) mechanism begins \citep{vdm2002}.There are some differences between these abundances and the ones listed in \cite{vd2010litio}, due to recomputation of some evolutionary tracks. There is a discontinuity in the lithium abundances as a function of the mass, that can be seen from Fig.~\ref{f1} and Table~\ref{t2}: for increasing initial mass, the lithium abundances first decreases, reaching a minimum of A(Li)$\sim$1.9 at 5.5\msun, then increases again. The initial decrease is due to the increasing temperatures at the bottom of the convective envelopes, that imply a faster lithium production and destruction. For the larger masses, the higher luminosities provoke larger mass loss rate at the stage of lithium production, and the abundance increases again. Nevertheless, there is a second minimum in the abundance at a mass of 7\Msun, that is the first mass for which the lithium production occurs {\it before} the beginning of the thermal pulse phase. For M=7.5 and 8\Msun\ the mass loss is dominant and huge quantities of lithium are lost to the interstellar medium, finally raising the lithium in the ejecta to values such as A(Li)=4.39 in the 8\msun. We warn that these numbers are dramatically dependent on the adopted mass loss rate \citep[see, for a discussion][]{vd2010litio}, nevertheless the discontinuity at 7\msun\ can not be attributed to simple numerics, and comes out from the above physical explanation, so we keep it in the ejecta table. 

Fig.~\ref{f1} summarizes the Li--Na--O abundances in the ejecta of
super--AGB and AGB stars from Table~1 in Paper~II. The most striking
characteristic of these abundances is the very large Li and Na
abundances reached in the ejecta of the 8\msun\ star. As remarked
above \citep[see also][]{vd2010litio}, this result is very uncertain,
as it may be due to an overestimate of the mass loss rate during this
evolution. Values of lithium abundances much smaller both for the 8 and 7.5\msun\ stars are possible, without significantly affecting the sodium and oxygen
abundances. As we will further discuss below, new observations will
allow to put tighter constraints on the
lithium yields of these stars. 

In Fig.~\ref{f1} we also show the much smaller oxygen values (due to deep
extra--mixing) that Paper~II assumes to be present at the surface of
SG stars formed directly from the super--AGB ejecta, and thus having
very high helium content. The values plotted refer to the abundances in the ejecta but,
obviously, the lithium abundances in giants will be much smaller than those plotted since  lithium  will be strongly diluted, both by standard and deep mixing.
The abundances in the ejecta of the masses 5--6.5\msun\ that
should contribute to the SG are A(Li)$\sim$2.0. Consequently, the AGB
ejecta will be important for the final lithium content in the SG stars
only in case B of Table~\ref{t3} (see Sect. \ref{assumedBB}), namely, when the Li abundance
in the diluting pristine gas is as low as that shown in the atmosphere of normal population II stars: if the initial Li is A(Li)=2.6, it is difficult that abundances 0.6~dex smaller (a factor 4 smaller!) in the
ejecta may alter the Li content of the mixture in a significant way
different from ejecta with no lithium at all. 

\subsection{Assumptions for the Big Bang Lithium abundance}
\label{assumedBB}
The application of the dilution model for lithium abundances is not as straightforward as for the other elements. In fact, {\it we do not know observationally} the lithium abundance of the pristine gas. The standard Big--Bang nucleosynthesis (BBN), with the cosmological parameter $\eta$=N(baryons)/N(photons) constrained by the observations of the satellite WMAP, provides A(Li)=2.64$\pm$0.04 \citep{spergel2007}, or even higher, A(Li)= 2.72$\pm$0.05 when updated rates are taken into account for the $^3$He($\alpha,\gamma$)$^7$Li reaction \citep{cyburt2008}. The values computed from the analysis of spectra of population II dwarfs are  A(Li)=2.37$\pm$0.06 \citep{melendez2004}, or A(Li)=2.23$\pm$0.07 in the non--LTE analysis  by \cite{asplund2006}\footnote{We quote an eye--averaged value for the data in the iron range --2.5$<$[Fe/H]$<$--1.2. Actually, the Asplund et al. (2006) results show a clear trend with metallicity. The trend becomes a spread at [Fe/H]$<$-2.7 \citep{sbordone2010}.} We have to  deal with the abundances into play by considering all the possible explanations of this discrepancy \citep[see][for a recent review]{fields2011}. The possible choices are:
\begin{enumerate}
\item  we reject the standard BBN scenario \citep[for a review, see, e.g.][]{iocco2009} and assume that the abundances observed in pop.~II stars (and in the FG stars of GCs) are the primordial abundances. Thus we adopt A(Li)=2.20 or 2.30 as value for the pristine diluting gas;
\item  we accept the standard BBN, but assume that lithium has been subject to depletion before the cluster formation, by means of the reprocessing of the primordial gas in a first generation of massive, hot stars (population III) \citep{piau2006}. As in the previous case, we adopt A(Li)=2.20 or 2.3 in the FG stars and in the diluting FG gas.
\item  we assume that Li is depleted in the atmosphere of both FG and SG stars, as a consequence of phenomena such as diffusion, gravity waves, rotational mixing, or any combination of these\footnote{The apparent lack of slope in the Spite plateau is an observational constraint that models can achieve by combining diffusion with some form of turbulence at the bottom of the atmospheric convective zone (Richard et al. 2005; Korn et al. 2006, 2007; Piau 2008; Lind et al. 2009b). The effect of turbulence is still introduced in a parametric way in the models.}. In this case, we assume A(Li)=2.60 or 2.7 ---as representative of the BBN abundance--- for the pristine diluting gas. We also assume that the atmospheric effect in SG stars is the same that depletes lithium in the FG stars, and reduces by $\delta$A(Li)=--0.3 to --0.5~dex the abundance of the mixed gas that forms the SG stars. Obviously, a depletion by 0.3 or 0.5~dex have different weight  on the fit of observed data. 

In order to begin exploring the range of parameters, we assume an initial A(Li)$_{\rm pristine}$=2.2, or A(Li)$_{\rm pristine}$= 2.7 for M4, and A(Li)$_{\rm pristine}$=2.3, or A(Li)$_{\rm pristine}$=2.6  for the models of NGC 2808. This is also very naively motivated by the FG abundances listed by different authors for the FG stars in different clusters.
\end{enumerate}

 \begin{table}
\caption{Possible cases of the lithium dilution when forming Second Generation stars and }             
\label{yields}      
\begin{tabular}{c c ll}     
\hline       
     &    Processed ejecta   &    Dilution  & Pristine gas \\
     &&& Abundance  \\
\hline                                                                      
A         &      Super--AGB --  AGB      &   Prist & BB A(Li)=2.6, 2.7      \\
B         &      Super--AGB --  AGB      &  Prist  & Pop.II  A(Li)=2.2, 2.3 \\
C         &      matter with N(Li)=0       &  Prist  & BB  A(Li) not relevant      \\
D        &      ---A(Li) reduced              &  Prist & like case A \\ 
E        &       in  Super--AGB---          &   & like case B \\
F         &      Super--AGB -- AGB        &  gas lost &  \\ 
            &                                             & from  stars & N(Li)=0   \\
\hline                                                                                                       
\end{tabular}
\label{t3}
\end{table}

\begin{figure}
\vskip -35pt
\resizebox{1.\hsize}{!}{\includegraphics{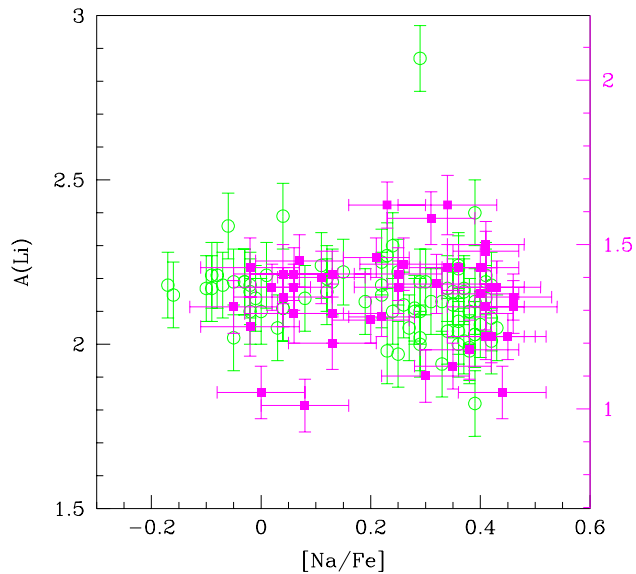}}
\vskip -55pt
\caption{The lithium versus sodium data in the sample by Monaco et al. (2012)  for turnoff -- subgiant branch stars in M4 (green open circles with error bars, left scale) are compared with the data by D'Orazi \& Marino 2010 (scale on the right) for giant stars. The abundance in the giants is depleted by convective dilution with respect to the abundance in the dwarfs, we assume that this dilution is uniform, and compare the data by a simple shift of the scale of the A(Li) logarithmic abundances.  }
\label{m4comp} 
\end{figure}

\begin{figure}
\resizebox{.48\hsize}{!}{\includegraphics{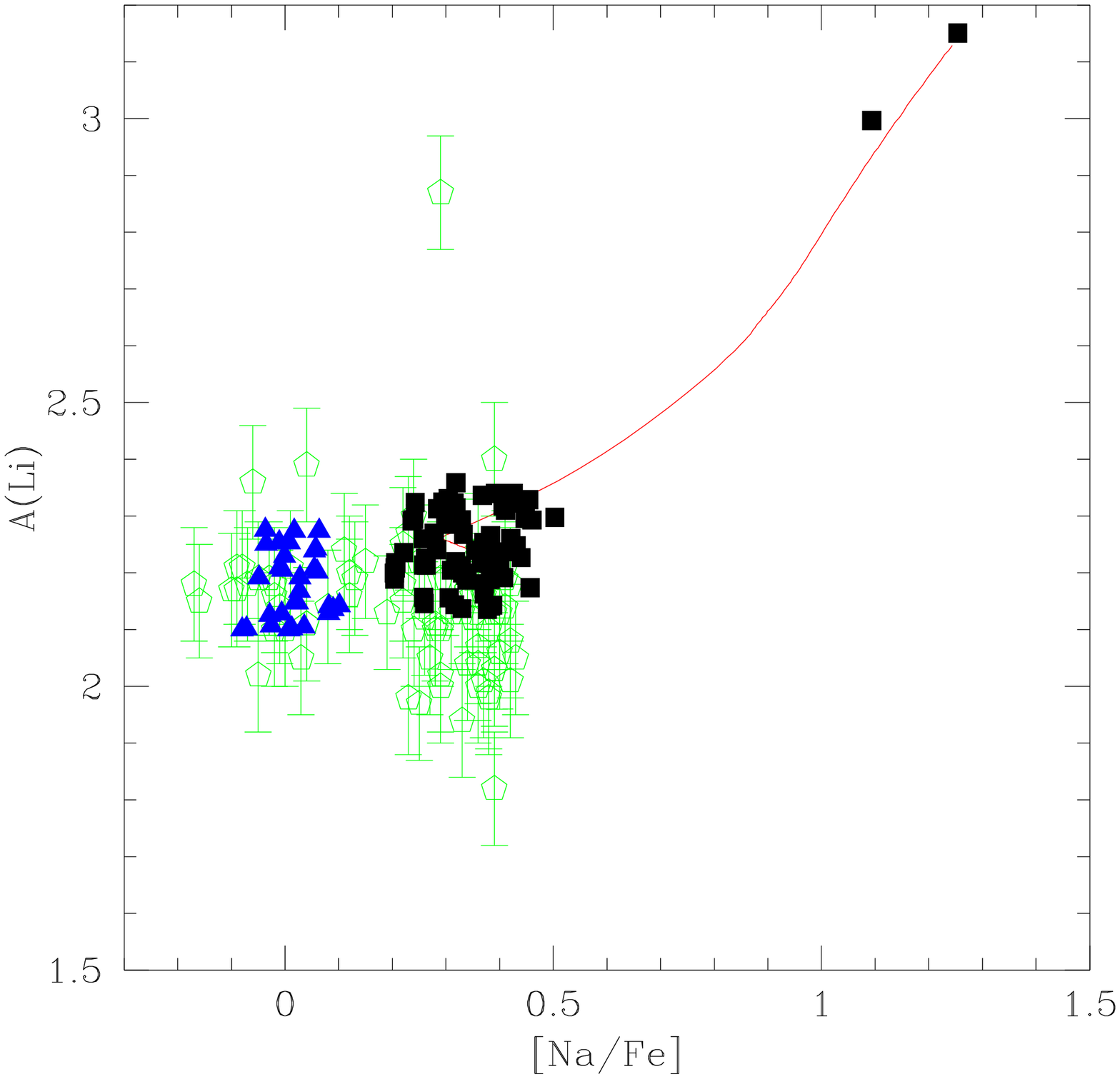}}
\resizebox{.48\hsize}{!}{\includegraphics{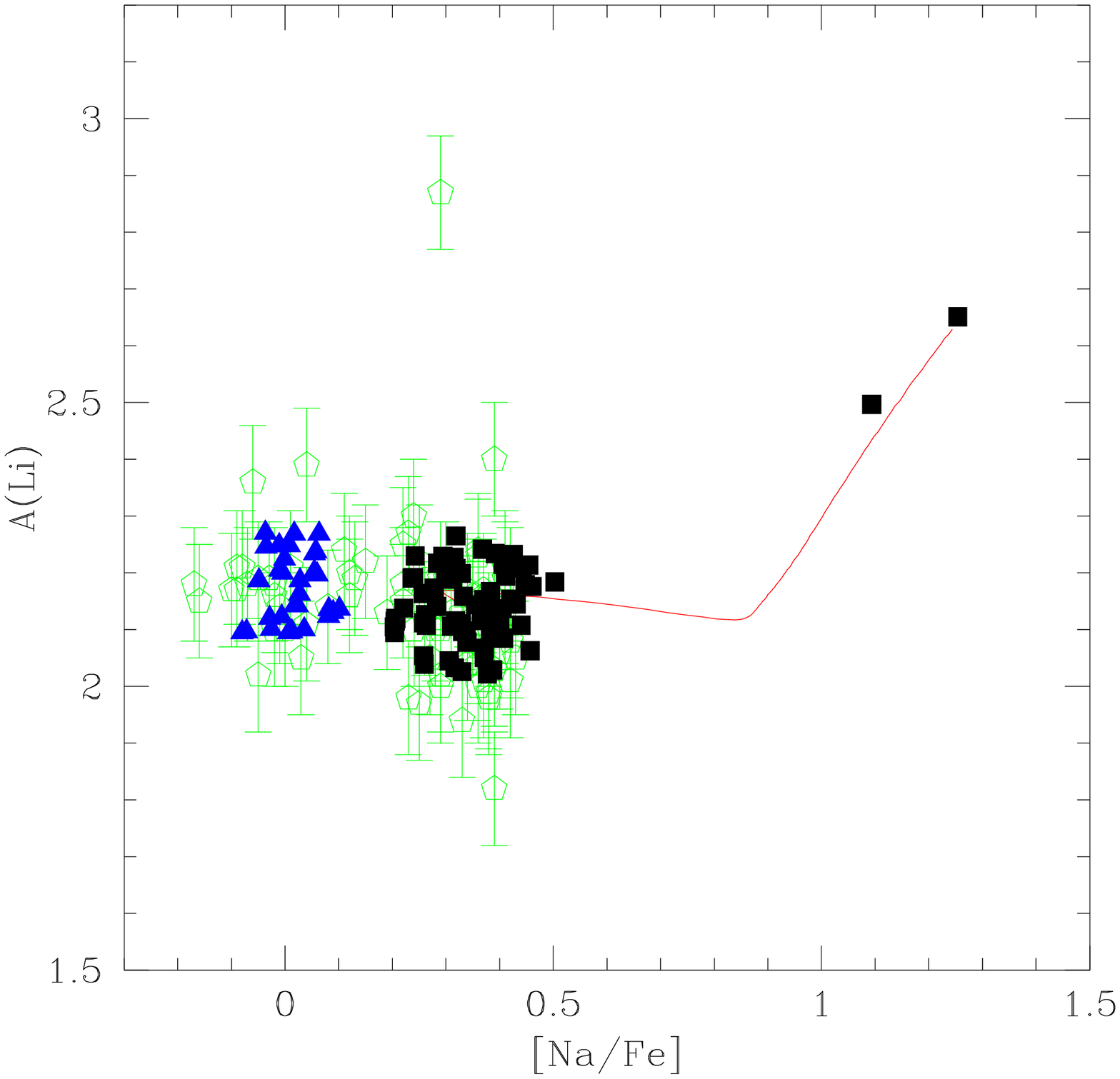}}

\vskip -25pt
\caption{The sample by Monaco et al. (2011)  for turnoff -- subgiant branch stars in M4 (open pentagons with error bars) are compared with the results of the simulation M4--0 (from Table~2 by D'Ercole et al. (2012)). Blue triangles are the first generation points, black squares are the SG points. The left panel shows case B (A(Li)=2.2 in the pristine diluting gas), the right panel represents case A (A(Li)=2.7). The red line in both panels represents the gas composition along the evolution. The initial part of the curve (at large lithium and sodium abundances) is scarcely populated due to the fast evolution of the gas composition and to the low star formation rate assumed (see Table \ref{t1}). }
\label{m40} 
\end{figure}

\begin{figure*}

\resizebox{.19\hsize}{!}{\includegraphics{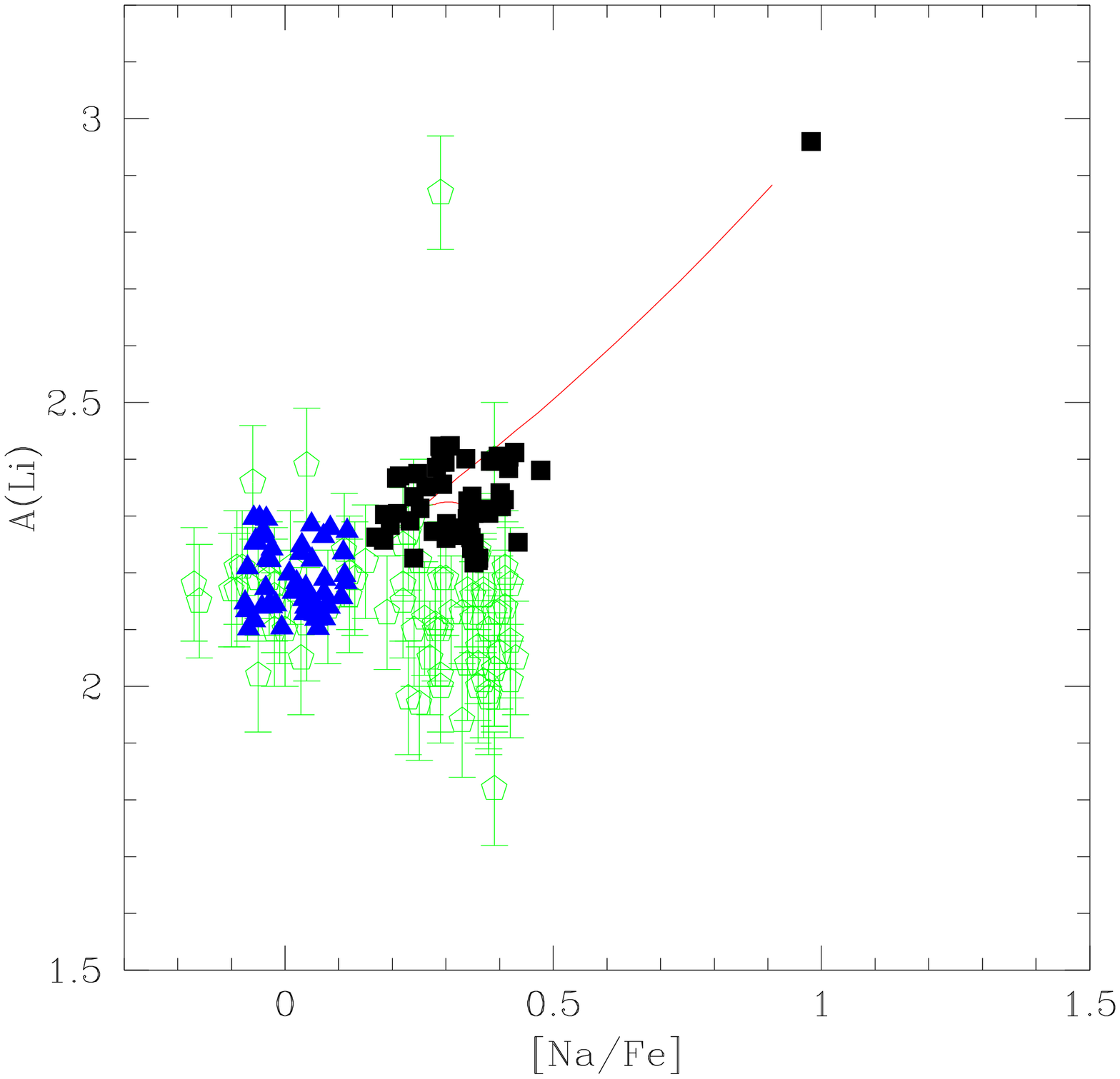}}
\resizebox{.19\hsize}{!}{\includegraphics{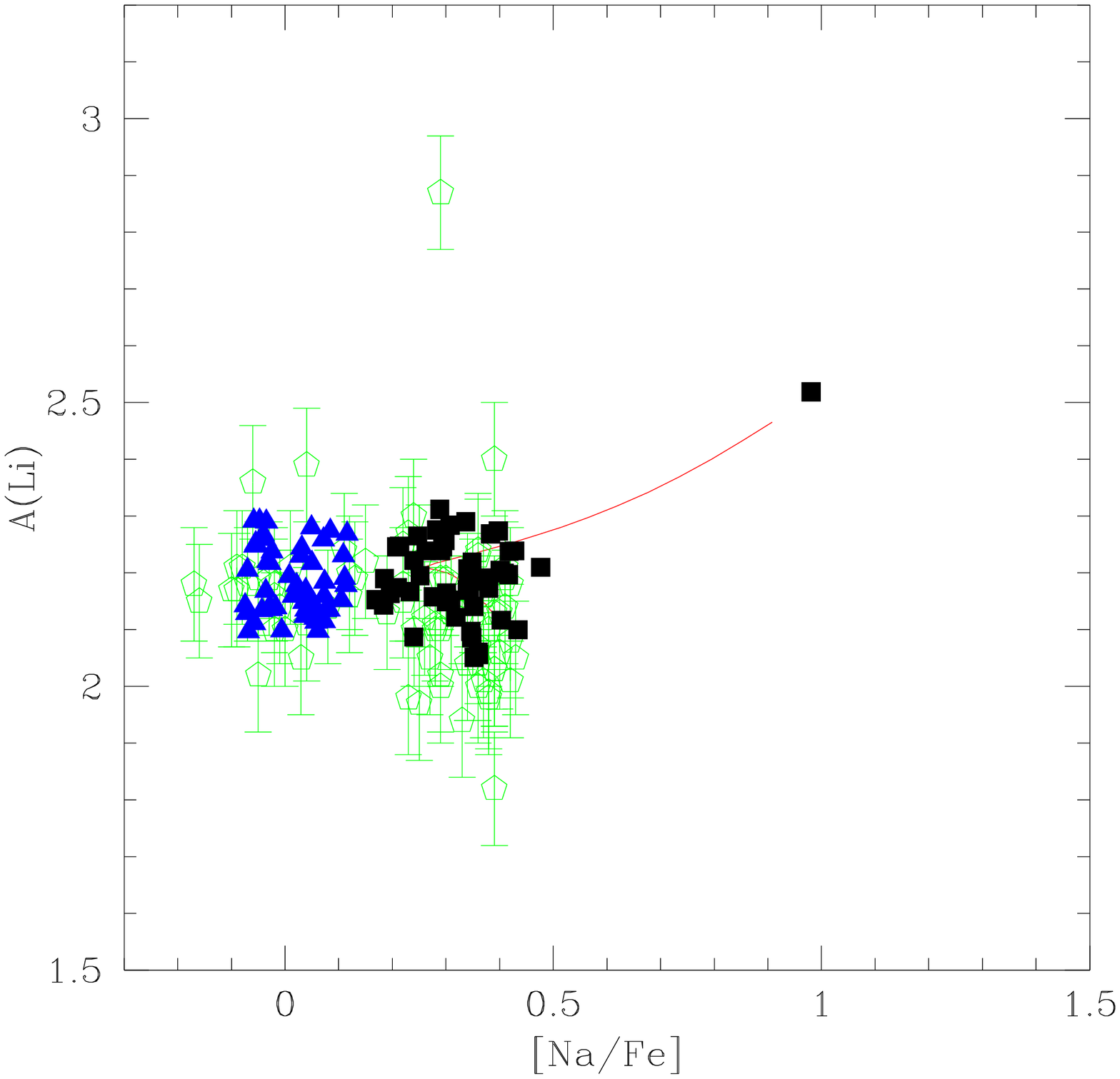}}
\resizebox{.19\hsize}{!}{\includegraphics{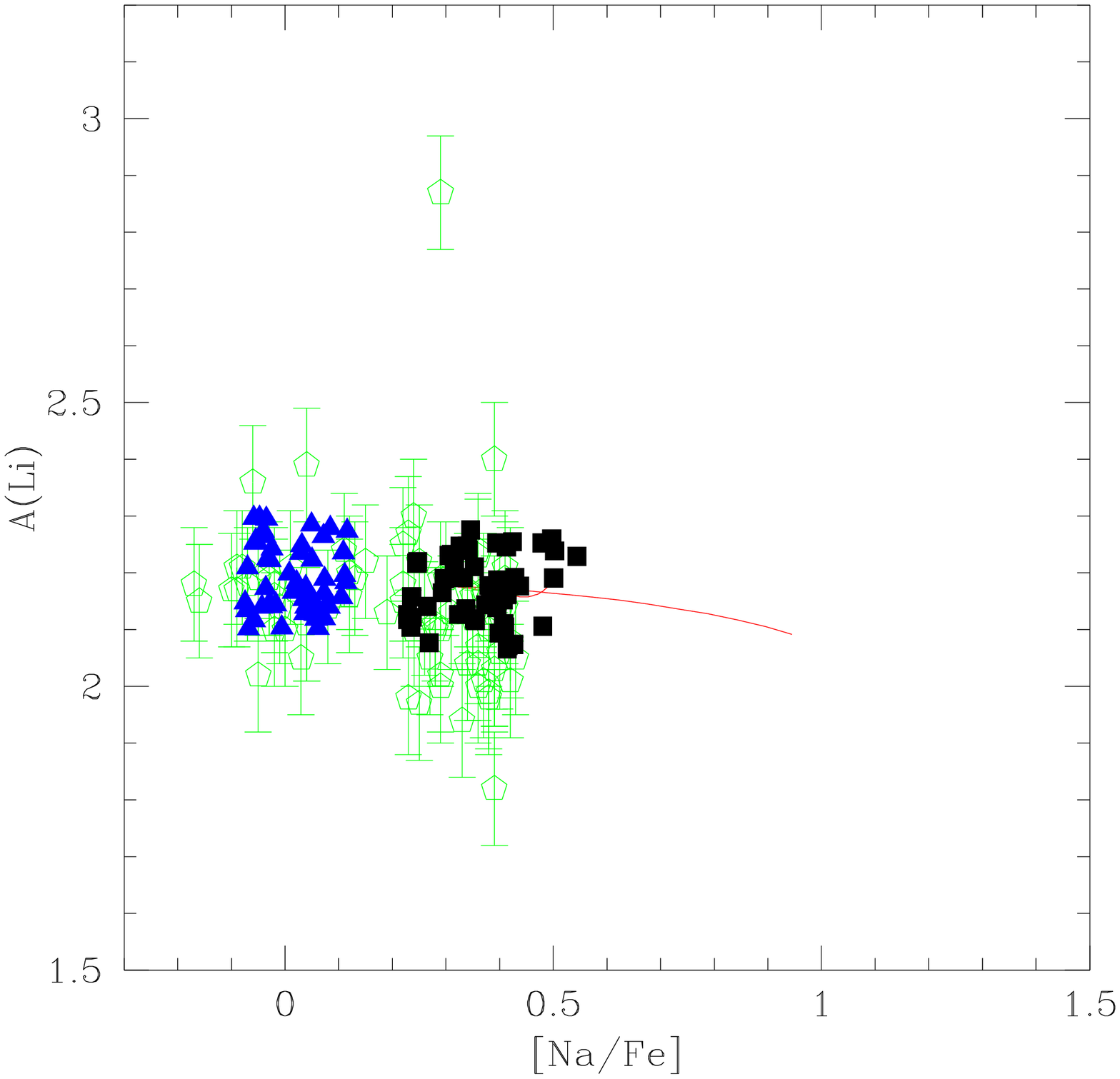}}
\resizebox{.19\hsize}{!}{\includegraphics{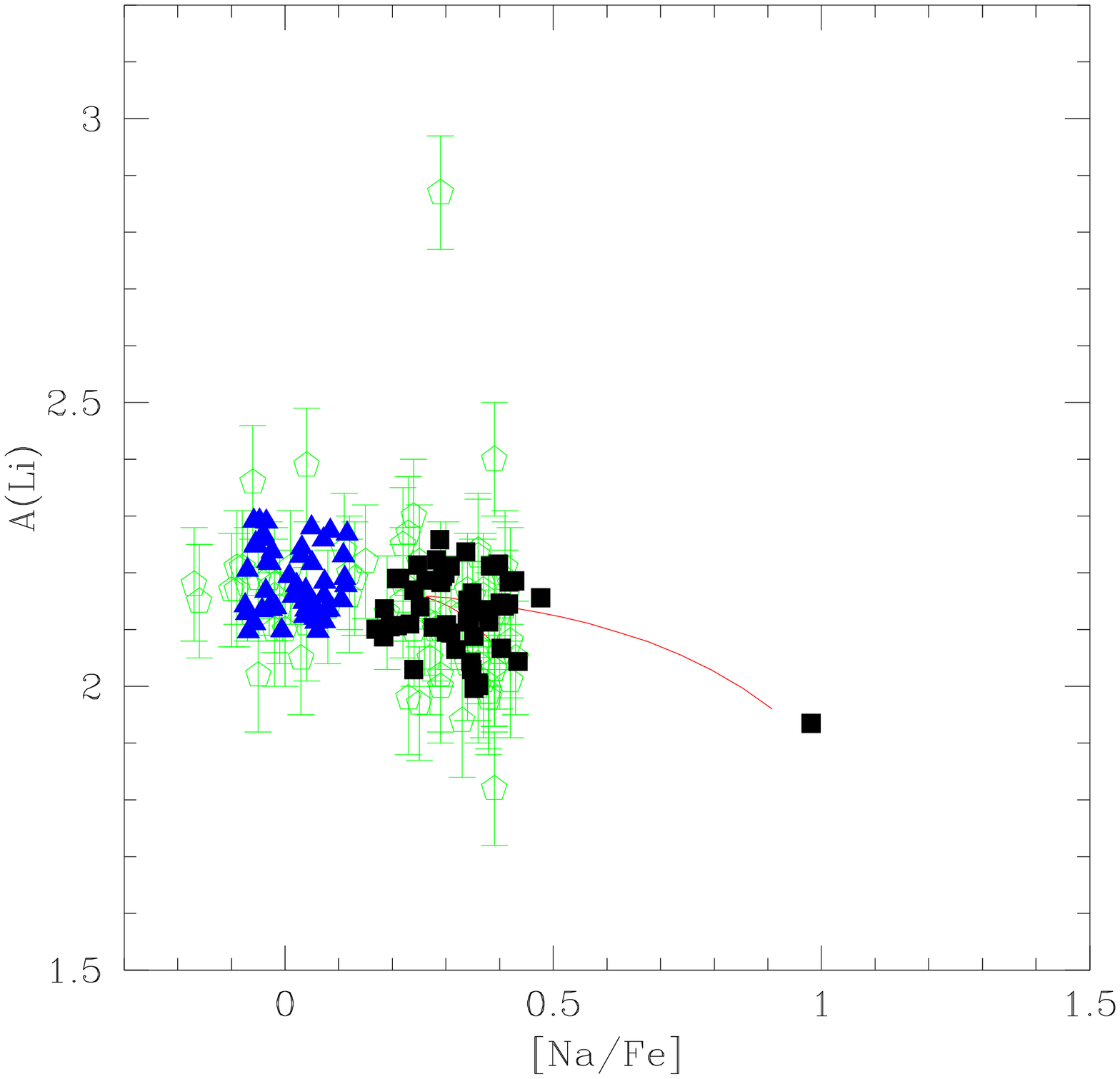}}
\resizebox{.19\hsize}{!}{\includegraphics{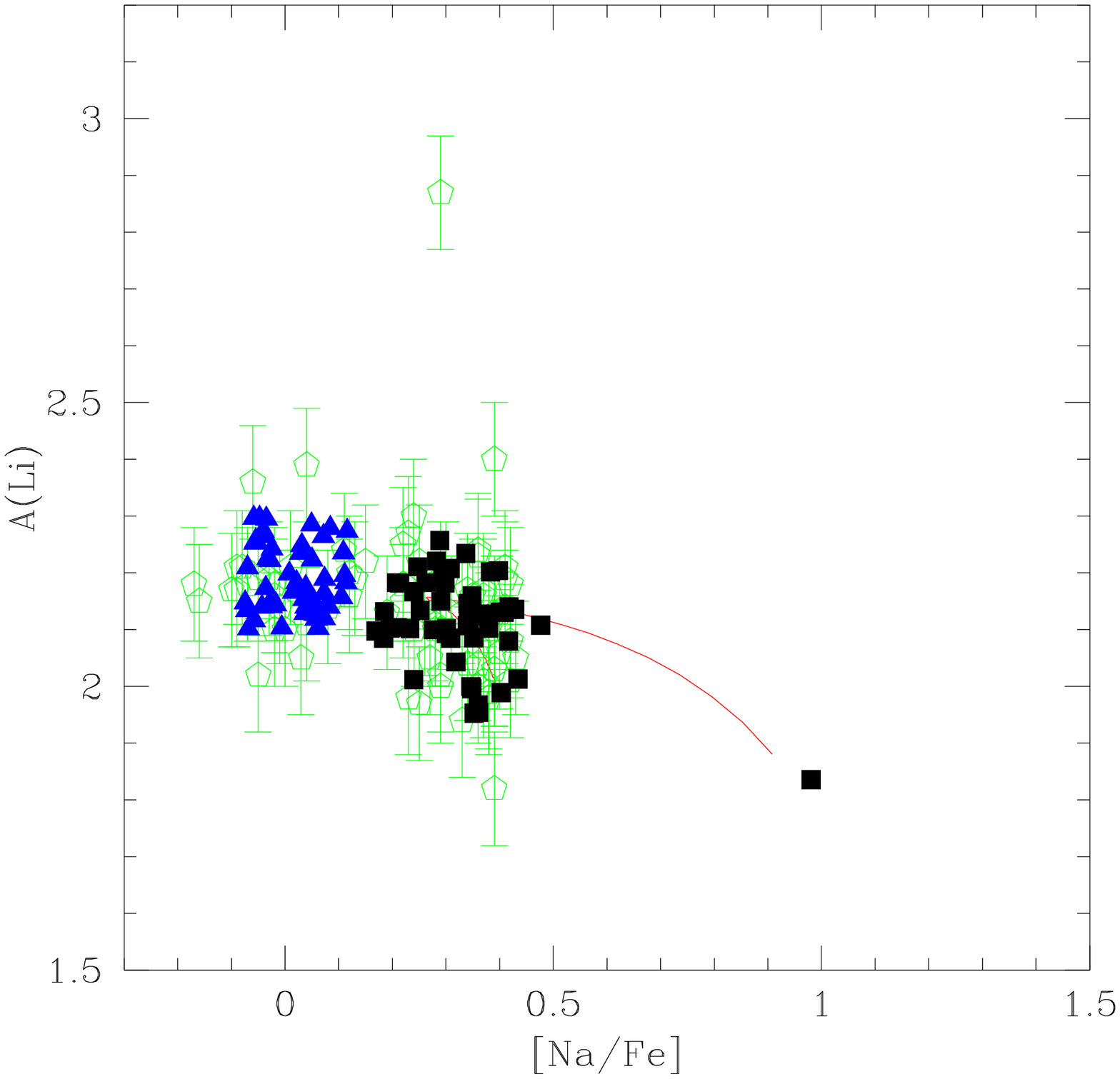}}
\vskip -25pt
\resizebox{.19\hsize}{!}{\includegraphics{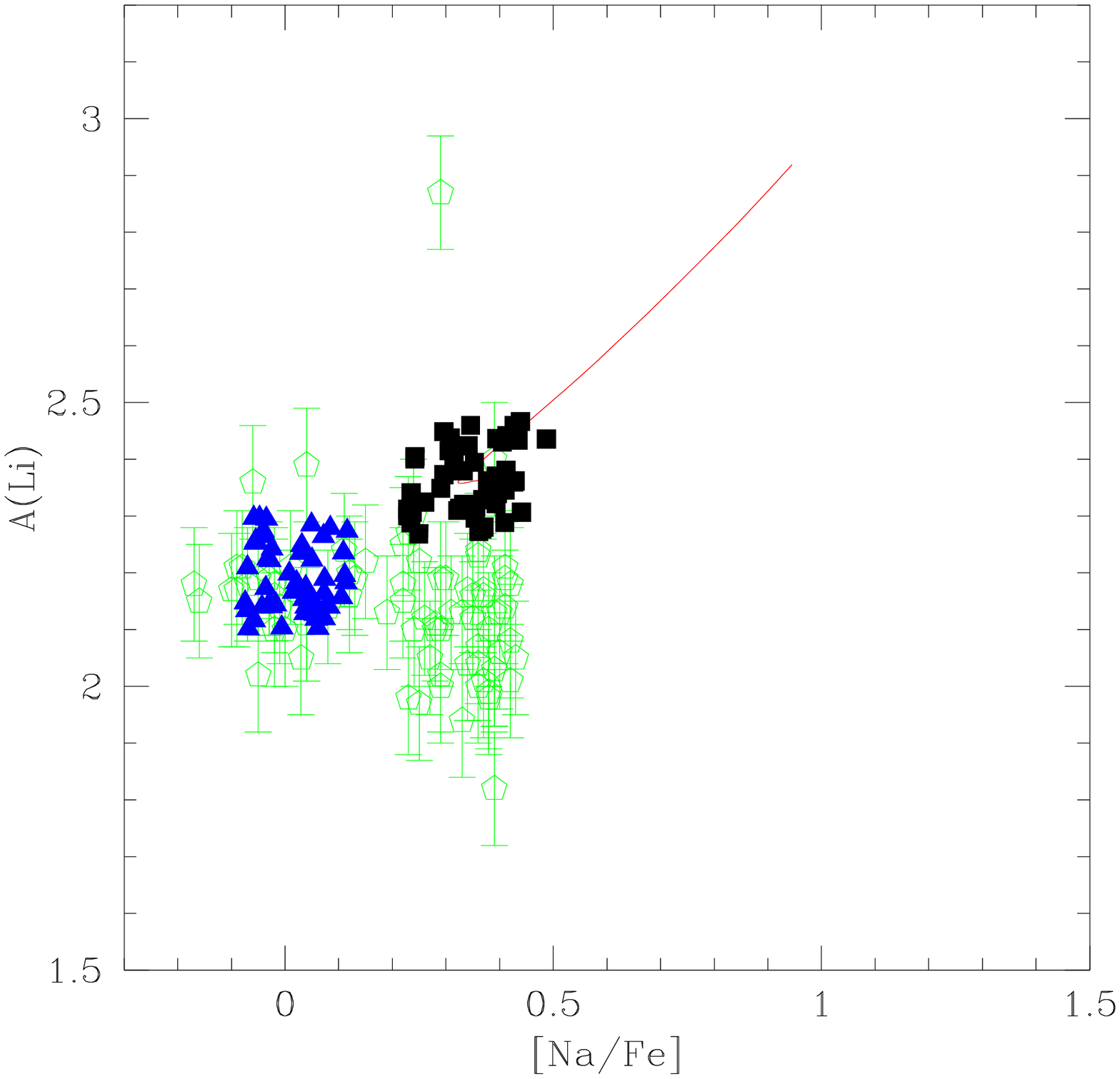}}
\resizebox{.19\hsize}{!}{\includegraphics{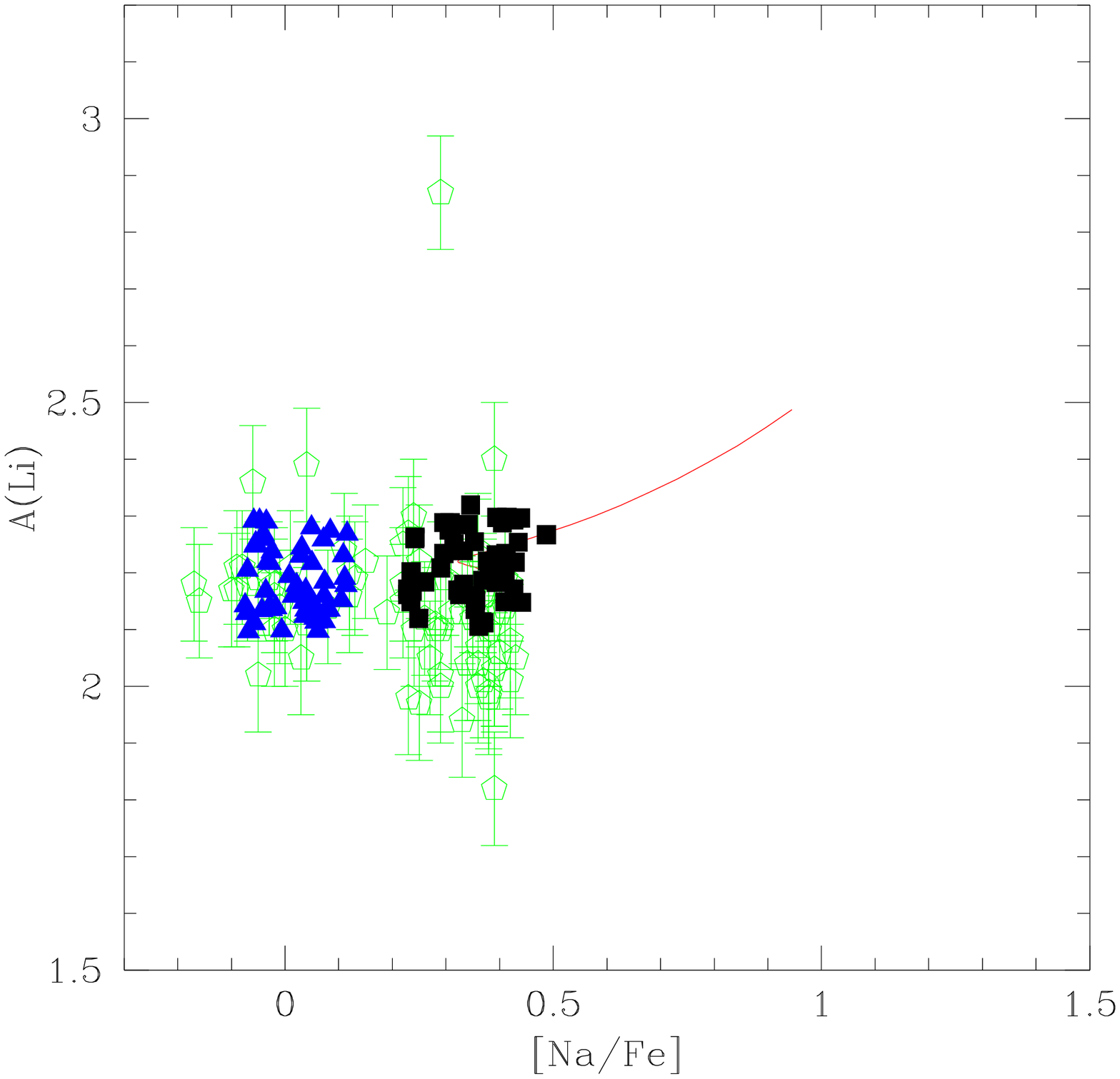}}
\resizebox{.19\hsize}{!}{\includegraphics{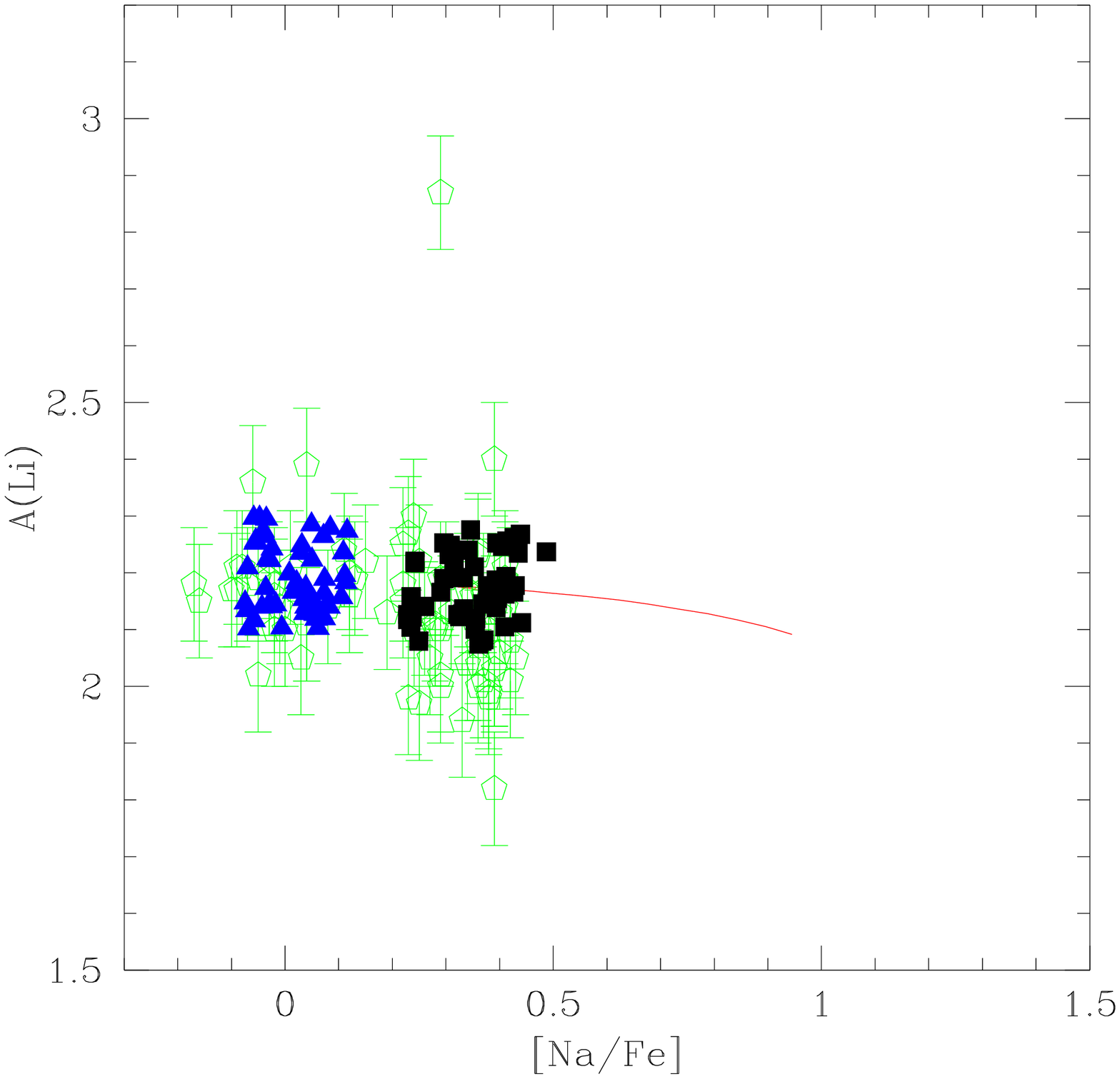}}
\resizebox{.19\hsize}{!}{\includegraphics{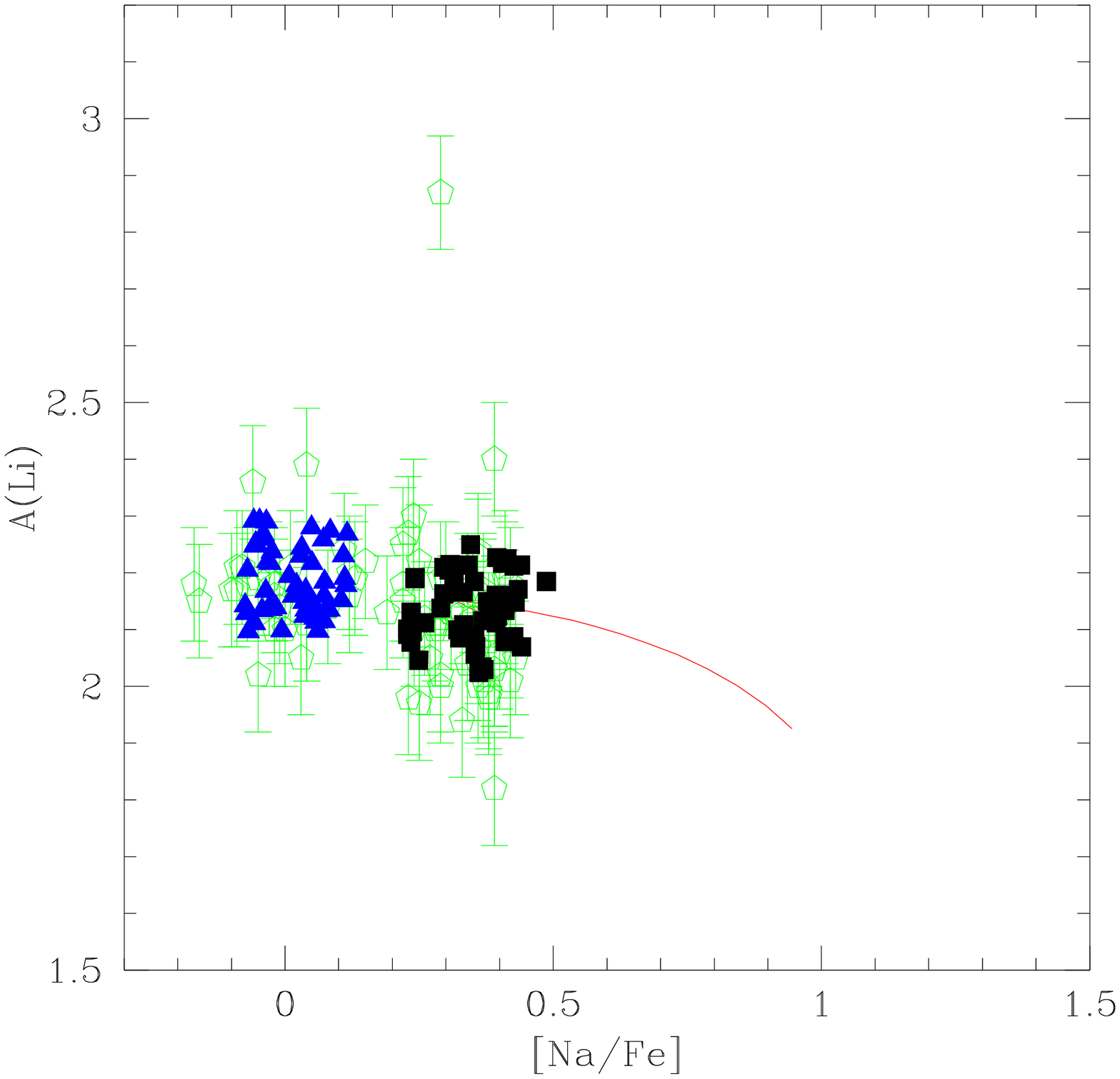}}
\resizebox{.19\hsize}{!}{\includegraphics{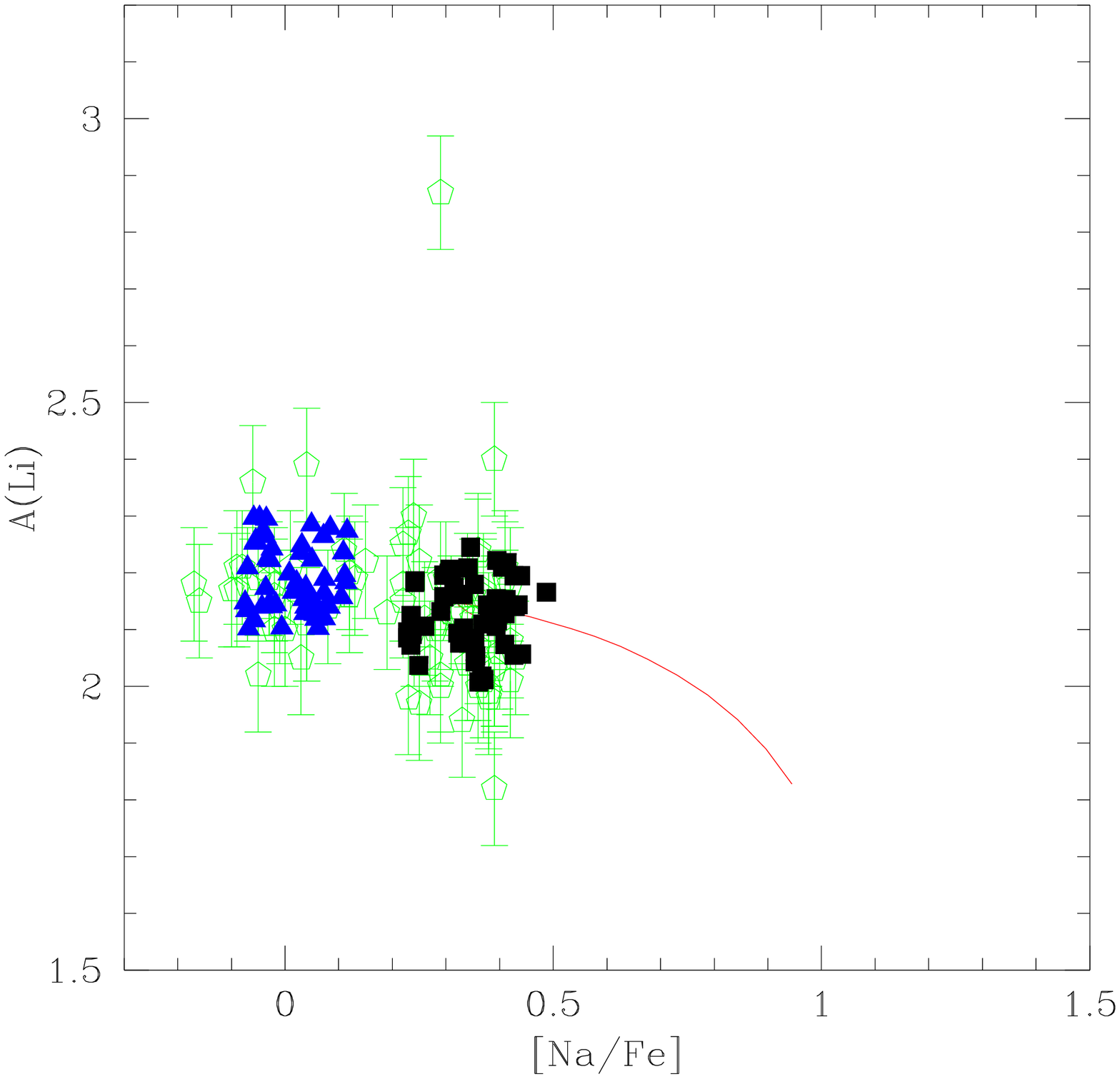}}
\vskip -25pt

\resizebox{.19\hsize}{!}{\includegraphics{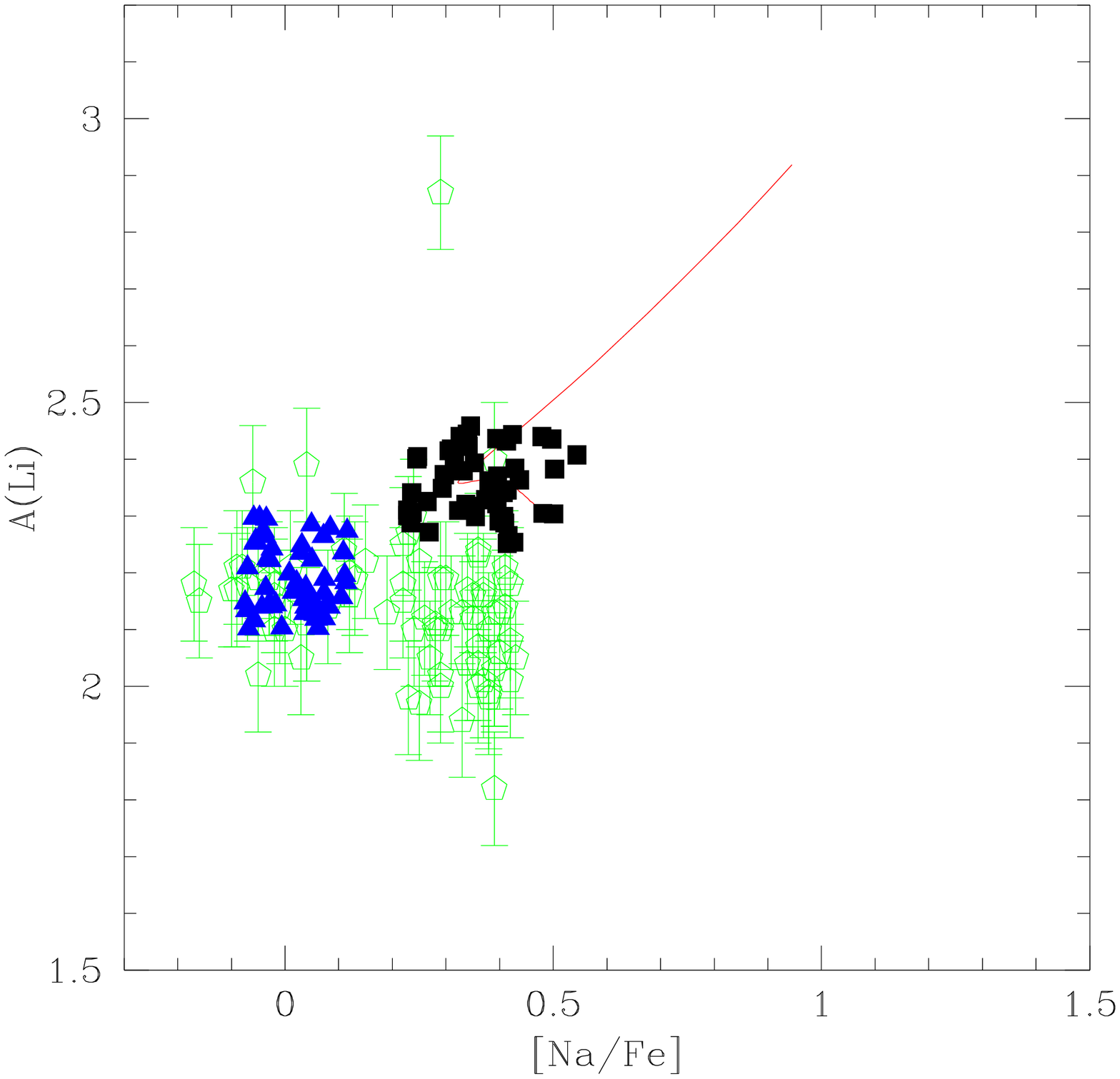}}
\resizebox{.19\hsize}{!}{\includegraphics{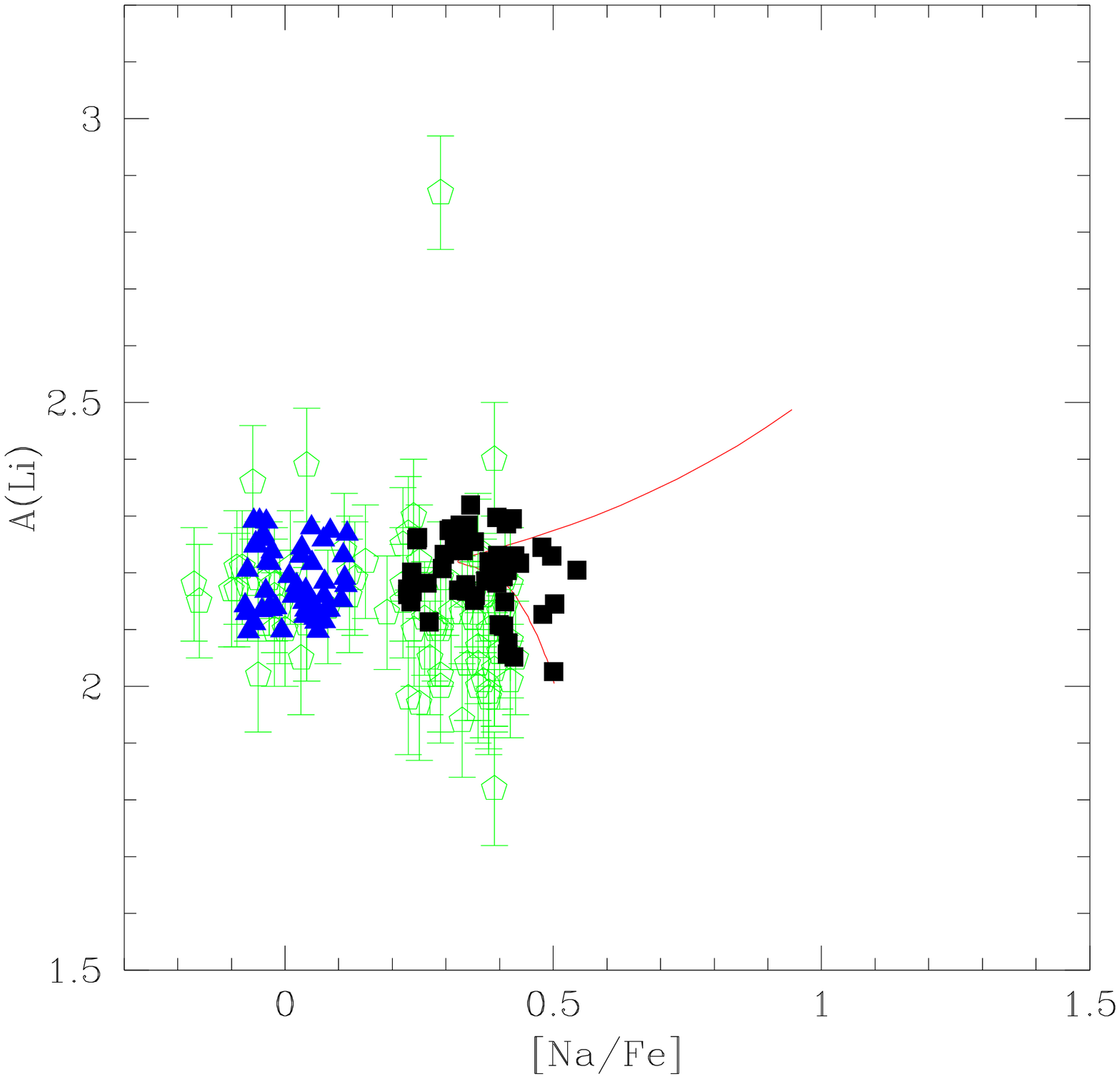}}
\resizebox{.19\hsize}{!}{\includegraphics{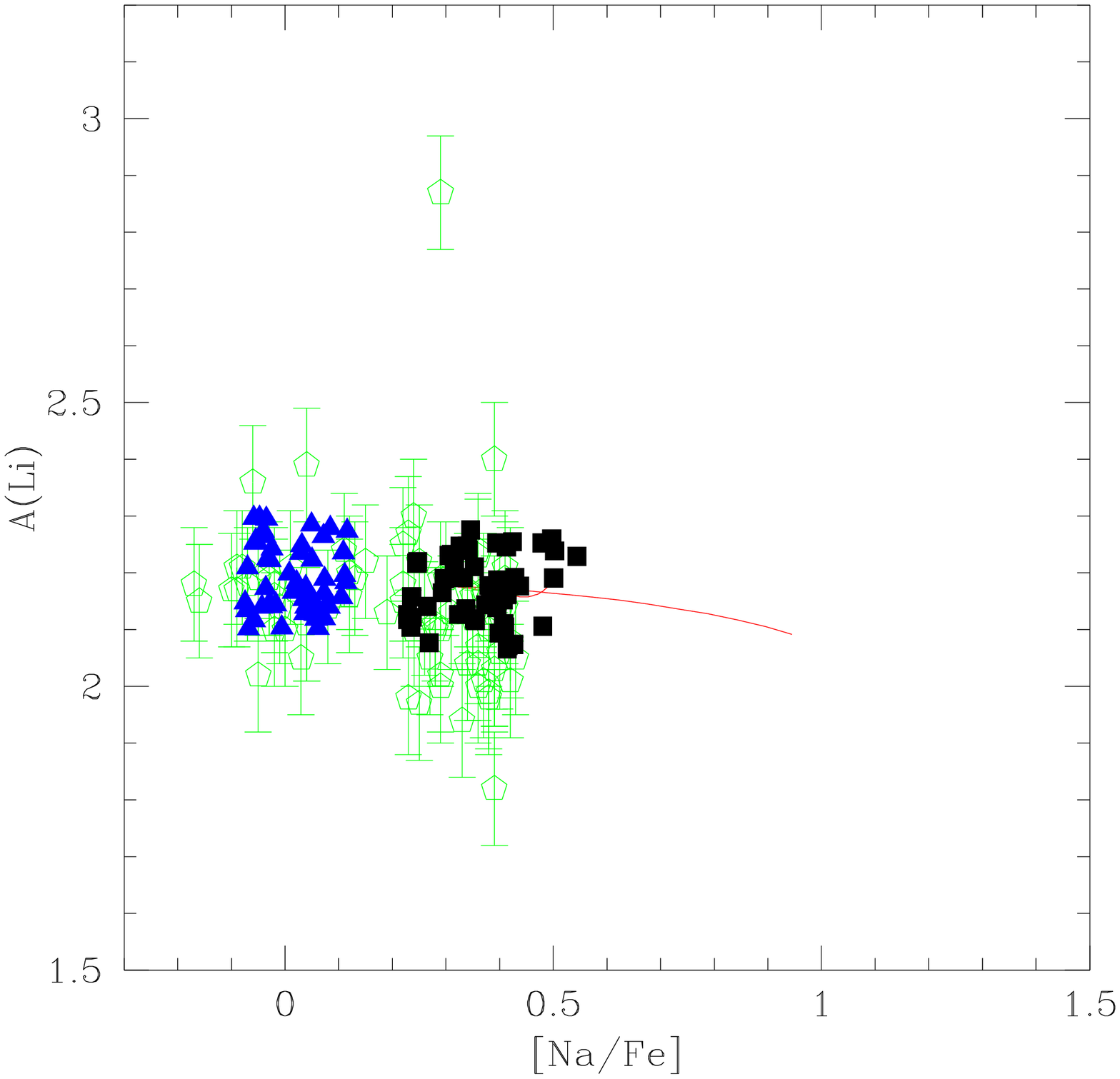}}
\resizebox{.19\hsize}{!}{\includegraphics{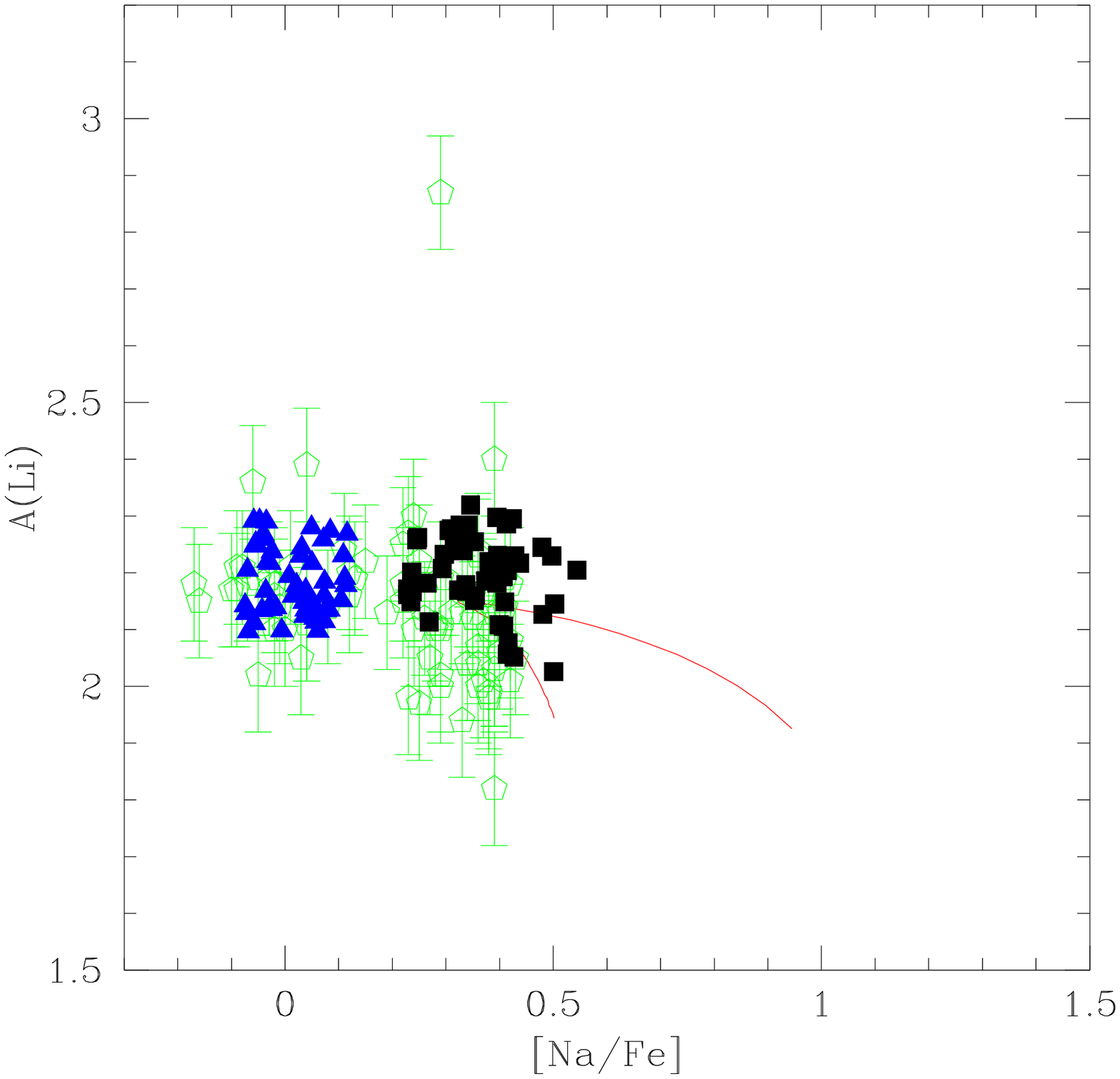}}
\resizebox{.19\hsize}{!}{\includegraphics{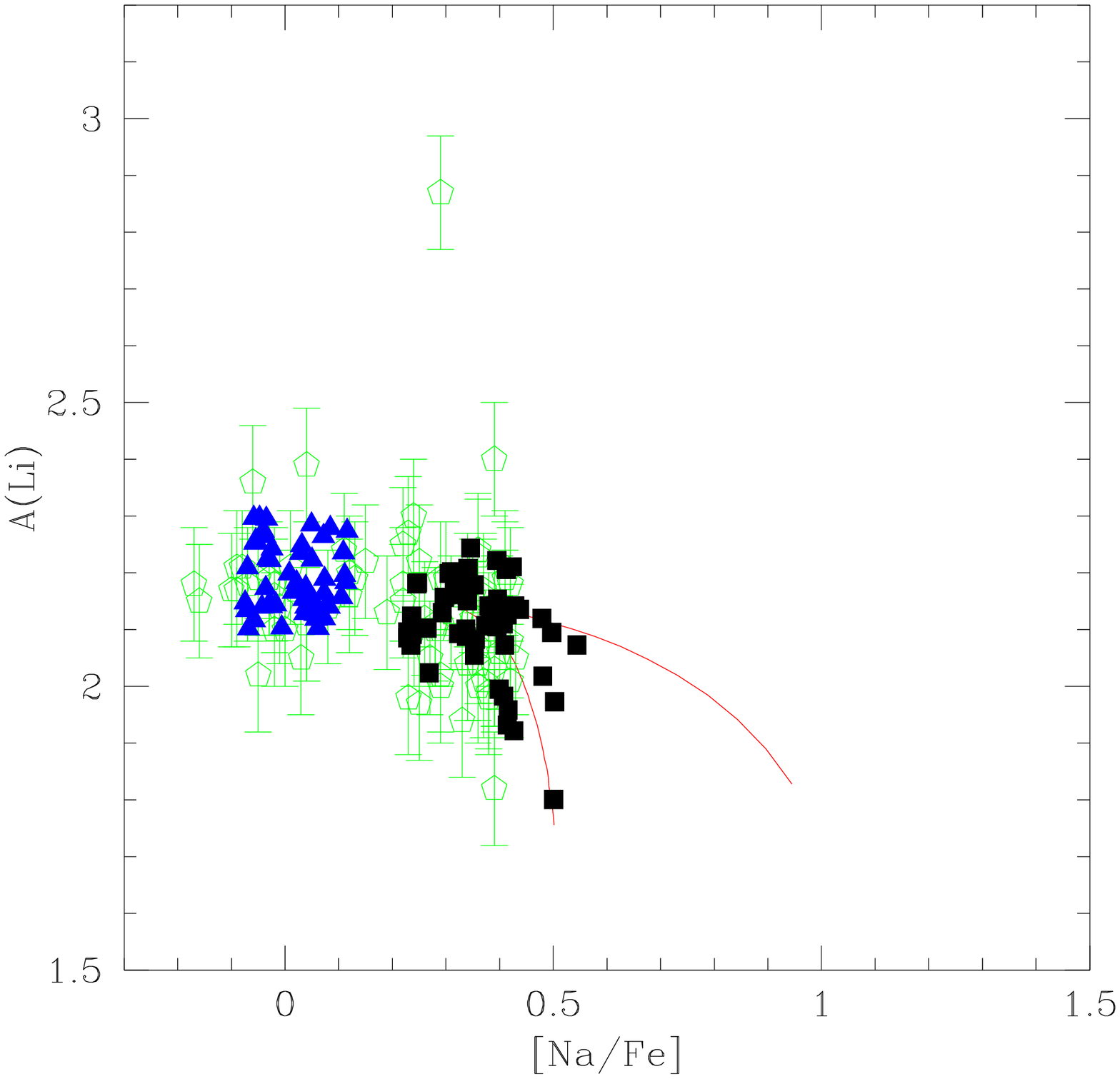}} 
\vskip -25pt
\resizebox{.19\hsize}{!}{\includegraphics{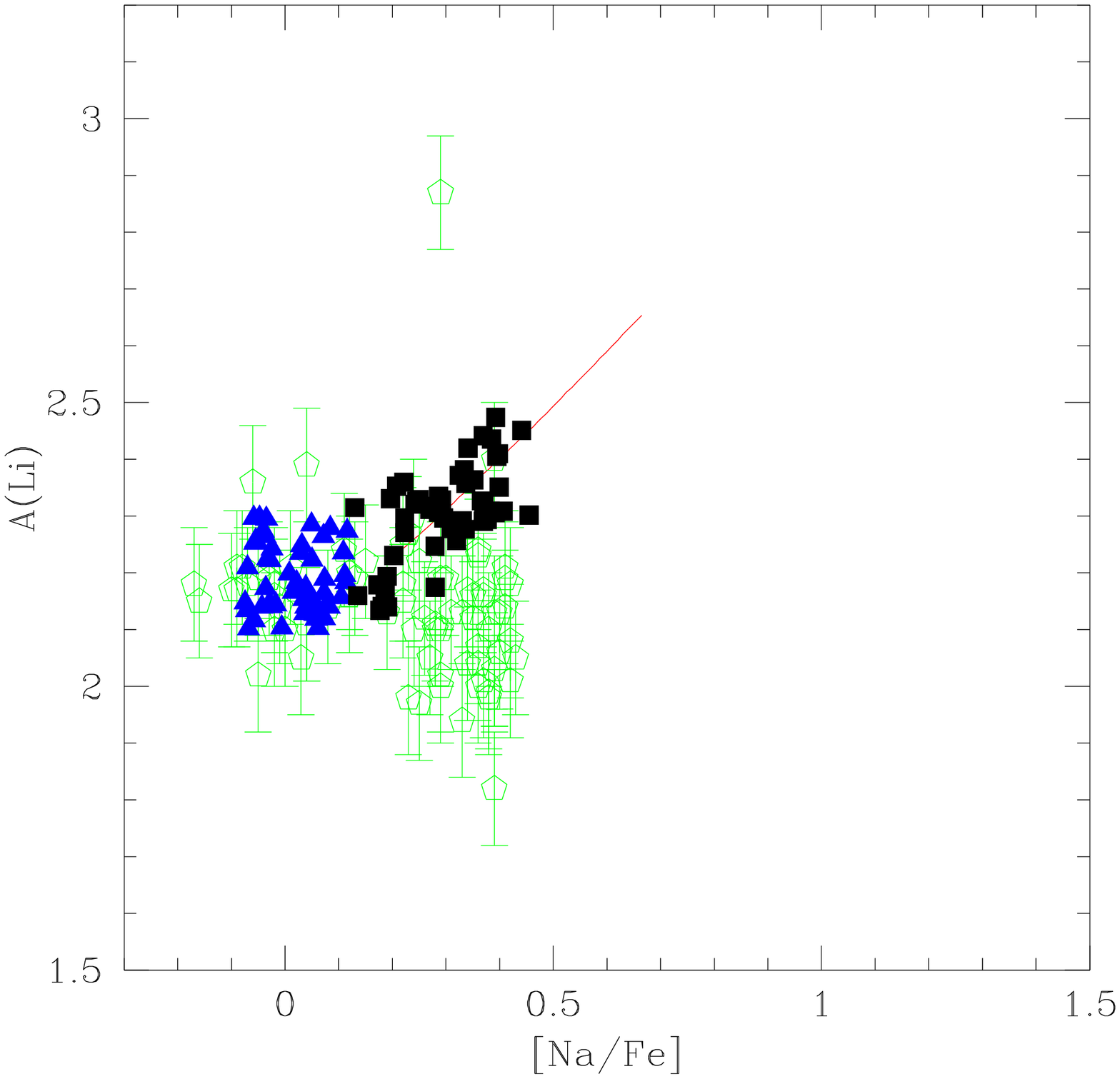}}
\resizebox{.19\hsize}{!}{\includegraphics{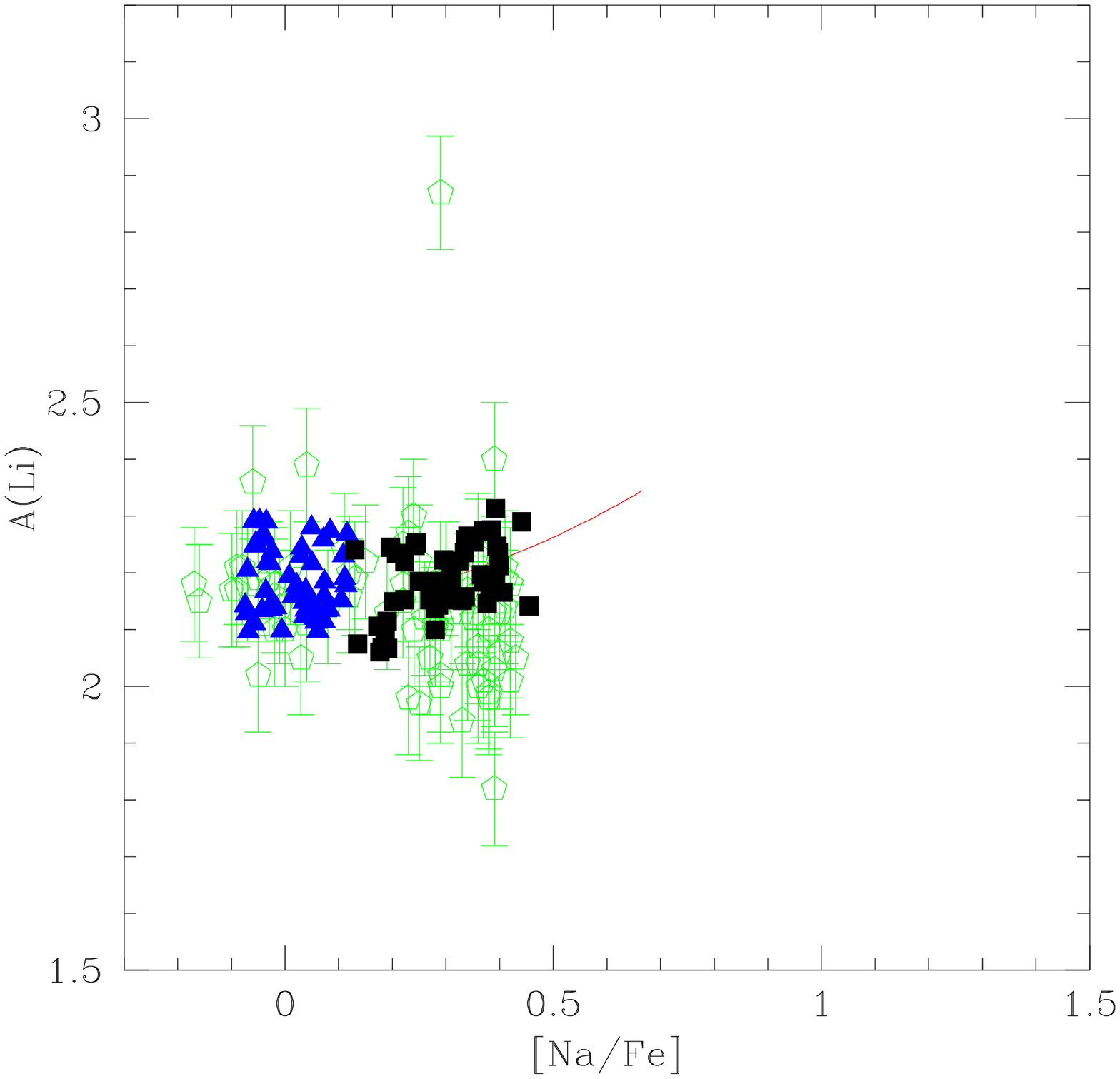}}
\resizebox{.19\hsize}{!}{\includegraphics{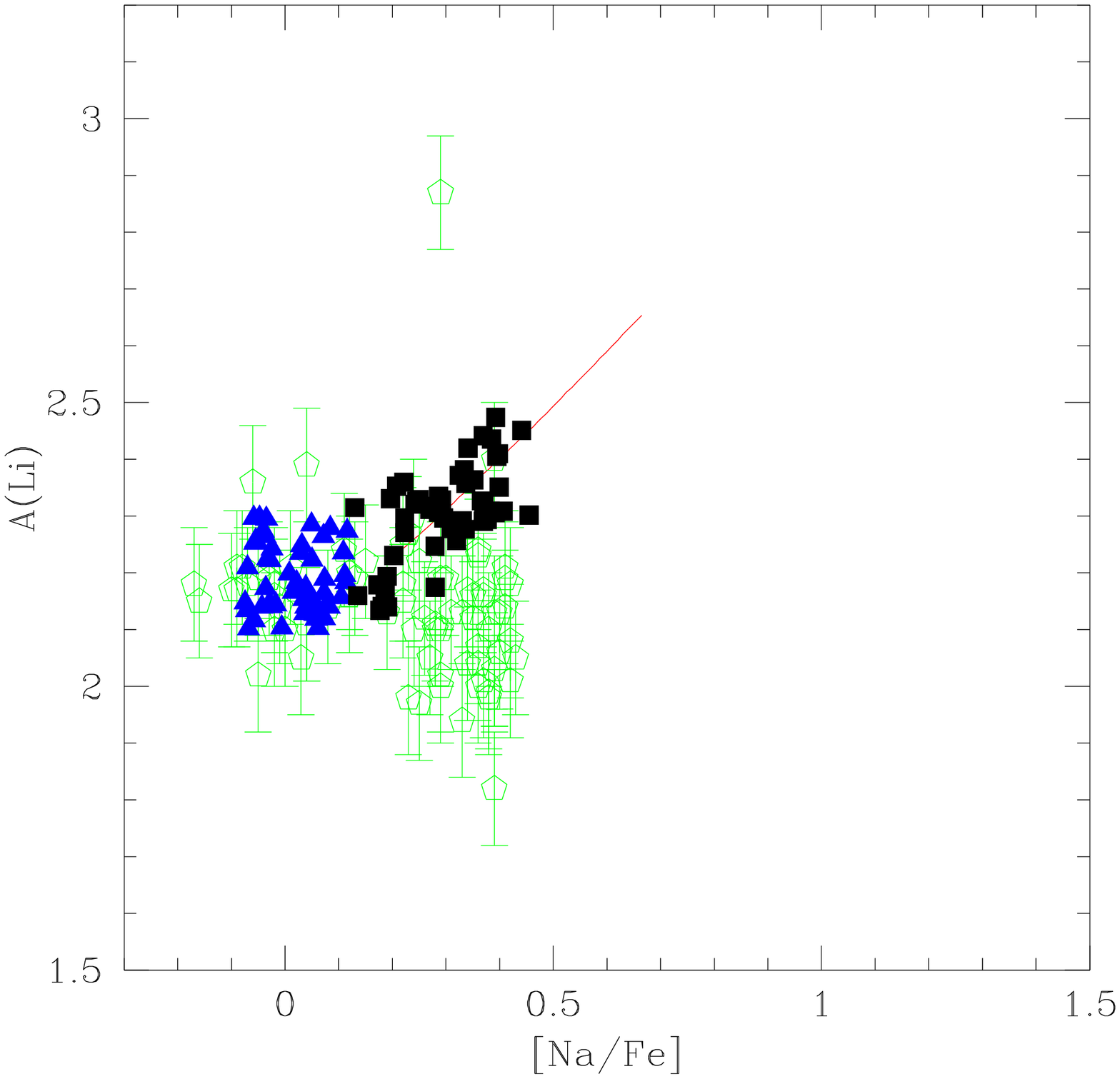}}
\resizebox{.19\hsize}{!}{\includegraphics{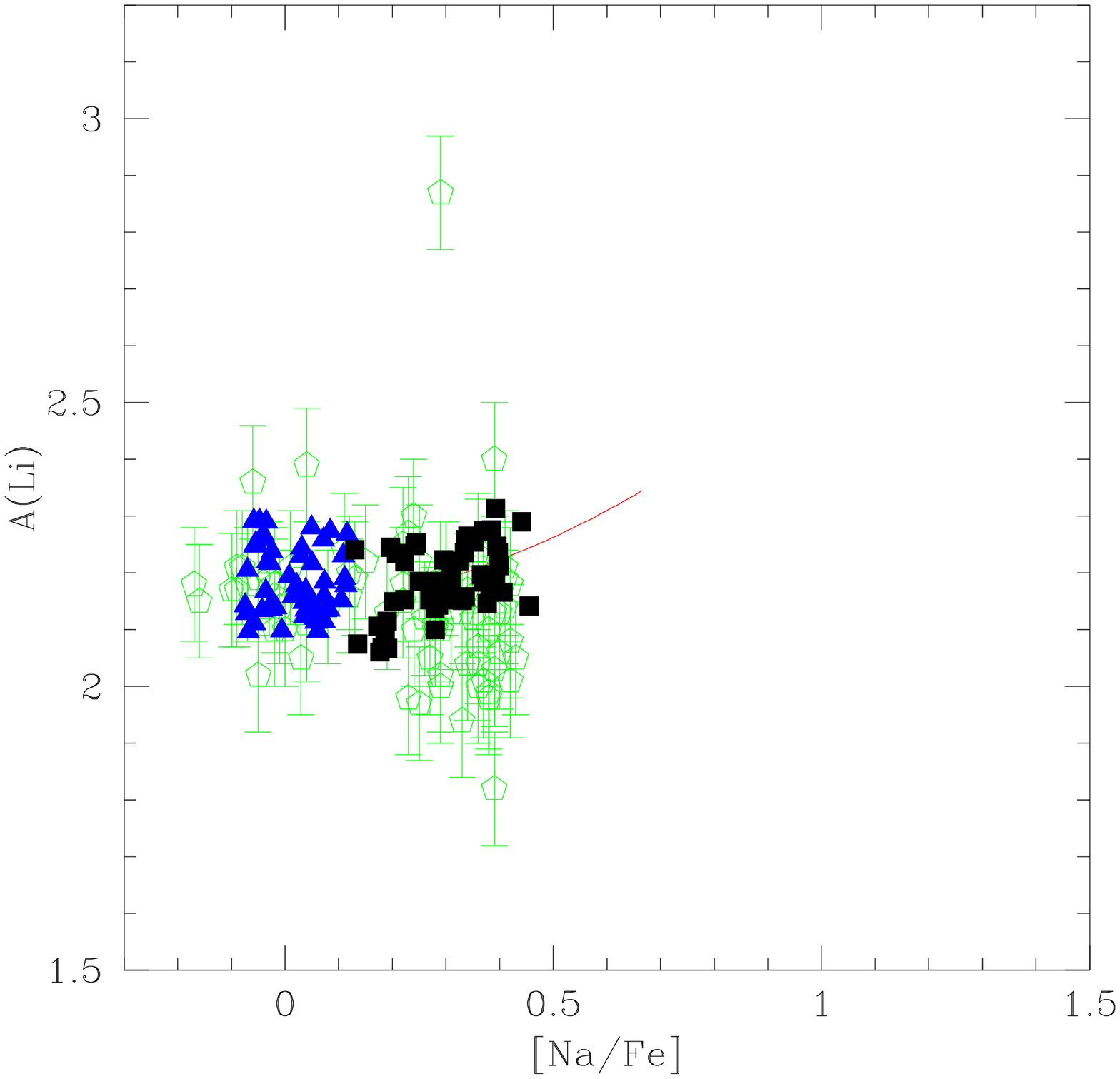}}
\resizebox{.19\hsize}{!}{\includegraphics{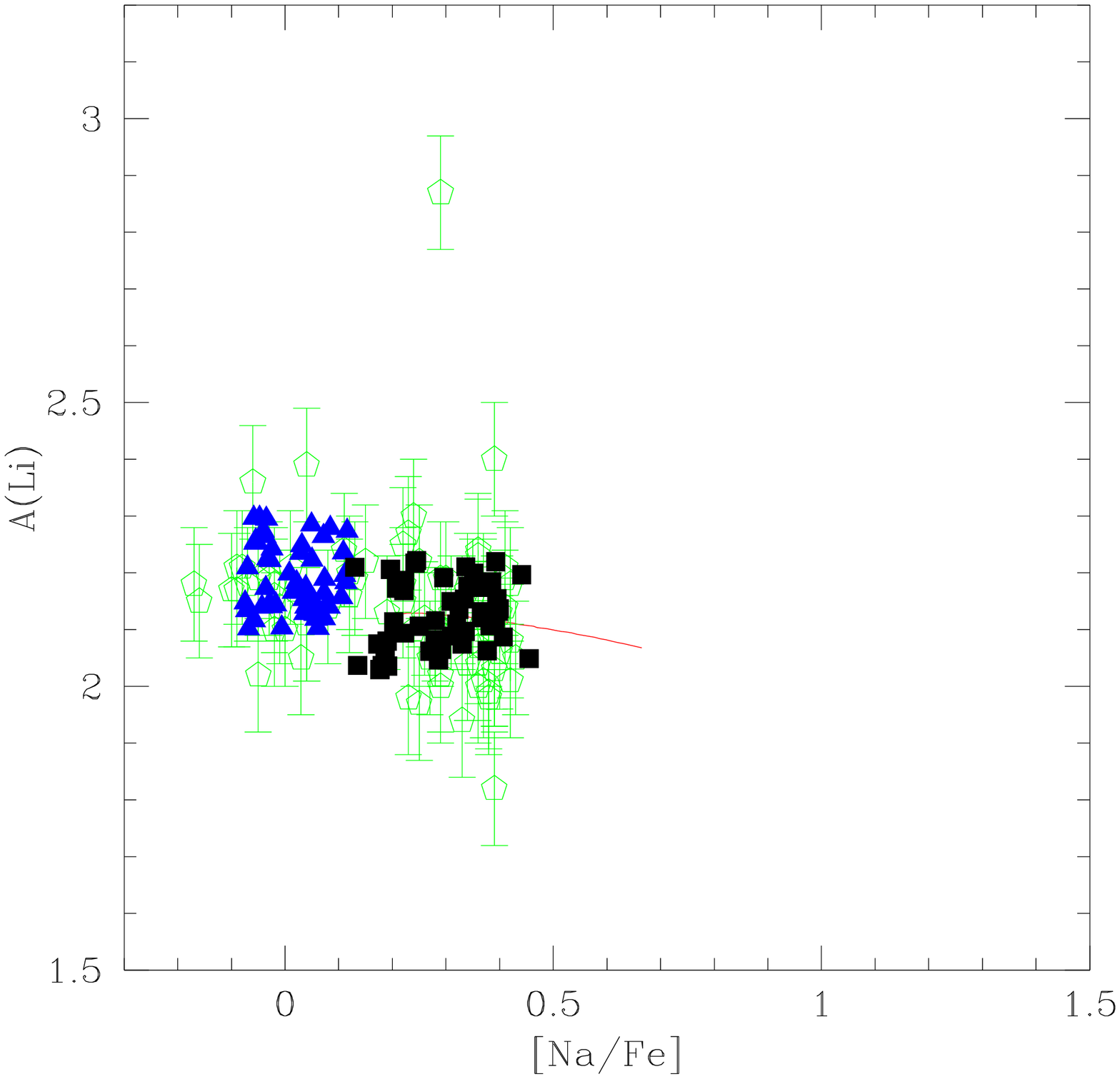}} 
\vskip -20pt
\caption{The data for M4 are compared with the simulations, from top to bottom: M4--2, M4--3, M4--4 and M4--5 (Symbols as in Fig.~\ref{m40}). 
From left to right, we show 1) case B (A(Li)=2.2 in the pristine diluting gas), 2) case A (A(Li)=2.7); following the two cases in which the lithium in the ejecta of the 7.5 and 8\msun\ are reduced to A(Li)=2.0, with A(Li)=2.2 (3) or 2.7 (4) in the diluting gas. Finally, we show the case in which we assume that the the polluting super--AGB matter has negligible lithium abundance (case C). In this last case, the content of the pristine gas (A(Li)=2.2 or A(Li)=2.7) is irrelevant.}
\label{m42} 
\end{figure*}

\subsection{Reference case}
Our standard case studies the SG stellar lithium abundances predicted  by some of the models we presented in  Paper~II and that well reproduce the abundance patterns in NGC~2808 and M~4. The lithium abundances in the ejecta of super--AGB and massive AGB are taken from Table~\ref{t2}, while all the other abundances are listed in Table~1 of Paper II. Table~\ref{t3} lists as cases A and B the two choices  discussed above concerning the lithium in the pristine gas. 

\subsection{Other cases}
We compare these ``standard" results with two other different possible combinations listed in Table~\ref{t3}: case C  in which the FG ejecta have no lithium at all (here the initial abundance of lithium in the pristine gas is irrelevant); cases D \& E, in which we reduce the yield of the 7.5 and 8\msun\ to much lower values A(Li)=2.0, and finally we consider a case F, in which AGB ejecta are diluted with gas having no lithium.

\section{The clusters with a mild O--Na anticorrelation}
\label{m4tot}
\subsection{M~4}
\label{m4}
In Fig.~\ref{m4comp} we show the recent data provided by \cite{monaco2012} for a sample of main sequence and subgiant stars in M~4, and compare them to the data for giants in the same clusters, given by \cite{dorazi2010}. The giants data (scale on the right) are shifted to take into account the convective dilution during the evolution. We
notice that the two samples provide a slightly different Li--Na trend:
the \cite{monaco2012} sample shows a slight negative slope (the
authors show that the difference in A(Li) between the low sodium (FG)
sample and the high sodium (SG) sample is $\sim$0.1~dex), while the
\cite{dorazi2010} data simply show a larger scatter in the higher
sodium data. The sodium zero point also differs  by $\sim$0.1~dex
between the two sets. As we will compare the models results with the
\cite{monaco2012} sample, the slight trend Li--Na is a significant ingredient of the comparison and affects considerably our interpretation. 

In Paper~I we provided a  ``standard" model to describe the ``short" O--Na anticorrelation of clusters like M~4, but in Paper II we have shown another class of viable models in which SG stars are formed from super--AGB ejecta strongly diluted with pristine matter. Fig.~\ref{m40} shows the results of the standard simulation (M4--0) of Paper~II, in which the bulk of accretion of pristine matter and star formation of the SG stars occurs during the evolution of the masses from $\sim$6.5 to $\sim$5\msun.  
 In this figure and in the following ones, we display as (blue) triangles the FG stars, and as (black) squares the SG stars. A stochastic error in the range 0--0.1 dex, comparable to the observational errors, is introduced, scattering the position of the stars, either with respect to the assumed FG abundances ([O/Fe]=0.4, [Na/Fe]=0, 
A(Li)=2.2) for the triangles, or with respect to the red line, representing the path of the gas composition during the SG formation, starting from top right and ending to the bottom left. 
The lithium abundance of the SG stars is very similar to that of the FG stars, if we are dealing with case B (A(Li)=2.2 or 2.3), because the lithium in the ejecta is very similar to what we assume to be the abundance in the pristine gas. On the contrary, if we deal with case A (A(Li)=2.6 or 2.7), the abundances in the ejecta are on average smaller than the pristine gas abundance, and the SG shows a slightly reduced lithium abundance, in full agreement with the data. \footnote{ In the standard case for M4 in Paper II we assume that the sodium yields for M4 are a factor two larger than those listed in Table 1 of that paper. This is done because neon is overabundant in the stars of this cluster, so that we should expect that also the sodium yield is larger (see the discussion in Paper I). For this reason, the two stars with very high lithium in this simulation have sodium as large as [Na/Fe]$\sim$1.3, a factor two larger than the abundance provided by the maximum super--AGB mass 8\msun\ ([Na/Fe]$\sim$1). }

In Fig.~\ref{m42} we show instead the results of the simulations made in Paper~II, in which the super--AGB have a dominant role for the formation of the SG. It is evident that the huge lithium abundances in the 7.5 and 8\Msun\ ejecta have led to an increase in the average SG star lithium abundance. In the leftmost column of Fig.~\ref{m42}, we show case B for cases M4-2, 3, 4 and 5 (from top to bottom). The average lithium in the SG is $\sim$0.15-0.2~dex larger than the FG abundance, while the observations show that it is $\sim$0.1~dex lower: this full factor two difference rules out case B\footnote{Notice that case B (and, with it, a Big Bang lithium abundance A(Li) as low as 2.2), is excluded ---if these models represent a correct reproduction of the SG formation in M4--- only thanks to Monaco et al. (2012) data, showing the mild anticorrelation Li--Na. It would not be excluded based on previous observations. Thus we remark here how crucial it is to get precise lithium determinations in multiple populations.}. The second column from left shows case A: here the situation is more ambiguous, and these models can not be excluded (see in particular case M4--2). In the third and fourth columns, we show respectively case E and case D --where we reduce the yield from the 7.5\msun and 8\msun to A(Li)=2.0:  both cases are acceptable, apart from the model M4--5 in the last row. Finally, the rightmost column shows the results obtained if we assume that the ejecta have no lithium, but maintain the sodium abundance of AGB--super~AGB ejecta in the polluting matter (case C). It is obvious that, in this case, assuming different Big Bang initial lithium is irrelevant.  The results show a very good agreement with the observations in all cases. 

We see then that the M~4 data {\it do not} allow to discriminate
between  the AGB scenario and other scenarios in which the polluting
matter is very sodium rich, but Li--free like in the FRMS model. Some
models of the AGB scenario reproduce well the M~4 Li--Na patterns,
but the Li in the super--AGB ejecta is not a necessary ingredient.

Fig.~\ref{m40} and \ref{m42} also show that the models predict the presence of a few stars with very large lithium abundance (see Sect.~\ref{2808} for a discussion of the difference in maximum lithium between case A and B). In M~4, Monaco et al. (2012) find that the star 37934 has A(Li)$\sim$2.9 ad relatively high sodium ([Na/Fe]=0.37). Could it  be born in the minute subsample of cluster stars formed directly from the super--AGB ejecta, just before the strong dilution with pristine gas that gave birth to most of the SG? 
Actually, we notice that the abundance of sodium predicted by the simulations is too large ([Na/Fe]$\sim$1), but a further exploration of the parameter space could be worth.
We also remark two consequences of this interpretation: 1) this star should be very helium rich. Its location at the left border of the main sequence turnoff stars (Figure 1 in Monaco et al. 2012) leaves this possibility open); 2) the HB of M~4 should include {\it a few hot blue tail stars}, representing the HB phase of these few very helium rich stars. 

Another cluster, NGC~6397 presents a case of a super--Li rich turnoff star \citep{koch2011}. Here A(Li)$\sim$4, close to our extreme predictions for the 8\msun\ super--AGB uncertain evolution. 
According to our models,  this super--Li rich star should also be characterized by a huge sodium abundance, but a preliminary study of its spectrum (private communication by K. Lind) appears to indicate a normale sodium abundance.

\subsection{NGC~6397}

The lithium abundances in the stars of NGC~6397 have been presented by \cite{lind2009}. We commented in Paper~II that NGC~6397 belongs to the clusters showing a mild O--Na anticorrelation \citep{carretta2009a}, and as such is a good candidate for a cluster in which the SG is formed by super--AGB ejecta strongly diluted with pristine gas. Unfortunately, we do not yet have models for the super--AGBs of metallicity as low as required to discuss more in detail this cluster (in particular, we expect smaller oxygen yields), but the interpretation of its data can be very similar to that presented for M~4 in Fig.~\ref{m42}. 
A strong dilution of the sodium rich, nitrogen rich and oxygen poor super--AGB ejecta may be an important factor to explain the problem of the Be abundance in two dwarfs of this cluster, analised by \cite{pasquiniber2004}. Both these stars have beryllium consistent with a spallation production of this element, a production that well describes the relation between beryllium and iron (or oxygen) in halo stars \citep{boesgaard1999}. Nevertheless, one of these stars is also oxygen depleted. \cite{pasquiniber2004} consider the possibility that this may be a SG star, and correctly notice that it should also be beryllium depleted, if the polluted matter included in its formation was hot--CNO processed. But, if strong dilution of the super--AGB ejecta with pristine gas has taken place, the beryllium abundance in this SG star may remain close to that of the pristine gas, similar to what happens for the lithium abundance.

\section{The clusters harboring an extreme SG}
\label{2808}

In this section we focus our attention on clusters that either have an
extreme O--Na anticorrelation (such as NGC~2808), or for which the
presence of a blue main sequence (MS) is evidence of the presence of a population with very
large helium content (such as NGC~2808 and \ocen), or have other signs
of an high-helium population,  like NGC~6752. In this latter cluster,
the horizontal branch shows a high-temperature tail, that puts into
evidence a high-helium content in a fraction of the SG stars
\citep{dantona2002}, and the main sequence too has some indication of
splitting \citep{monaco2010}. Although this cluster is probably not as
extreme as \ocen\ and NGC~2808, it also seems to harbor three MSs identified by means of UV colors
(Milone et al. 2012, in preparation). In addition
 it has an interesting set of O--Li abundances \citep{shen2010} that are worth of an analysis in terms of our models. 

\subsection{Predictions for the extreme populations: the blue MS of NGC~2808 and  \ocen}
Paper II reproduces  the O--Na anticorrelation and three different helium contents of the subpopulations in NGC~2808 by a two-stage process of SG formation, the first episode taking place from the pure super--AGB ejecta, leading to the very high helium stars populating the blue main sequence, the second episode occurring from ejecta diluted with pristine gas. As discussed, different models are possible for the timing and duration of the episode of pristine gas accretion, all consistent with the observed O--Na anticorrelation. 
In Fig.~\ref{f3} we now examine the predictions for the Li--Na and Li--O patterns  for the model NGC~2808-1 of Paper II, that reasonably represents both the O--Na anticorrelation and the probable helium difference between the FG and the extreme and intermediate SG displayed by NGC~2808. 
The SG points are 40\% of the total, following the choice made in Paper II.  
The number of points is proportional to the time spent along the line (as well as to the star formation rate)
so that very few stars have the extreme compositions that are linked to the short phase during which the pollutors are the most massive super--AGBs. 

\subsection{The extreme population}
Large lithium abundances are found in stars formed from pure super--AGB ejecta. 
We notice again the interesting role played by the value assumed for
the Big Bang (or pristine) lithium abundance, by comparing the left
and right panels, representing case B (A(Li)\pristine=2.3) and case A
(A(Li)\pristine=2.6). In fact, the {\it maximum} Li achieved (for a given model of SG formation) 
is larger in case B, where we
assume that the pristine gas has the same lithium abundance observed
in population II stars. In this case, the atmospheric
abundance coincides with the initial gas abundance, as no reduction is
ascribed to mechanisms that deplete it. If the pristine gas has a
larger, e.g. a standard Big Bang A(Li)=2.6, we implicitly assume that
depletion mechanisms are acting at the surface, to bring this value
down by $\sim$0.3~dex. These same mechanisms will apply also if the
lithium abundance in the star is the result of the super--AGB ejecta
high abundance. 

Thus, contrary to simple intuition, lithium in SG
stars (and in particular its maximum abundances, that are directly
related to star formation from pure ejecta) will be $\sim$0.3~dex {\it
  smaller} if the diluting gas has a {\it larger} abundance, {\it in
  the stars that form from pure ejecta}. This could not be noticed in
the bulk of results for M~4 (but is evident in the two stars with
extreme abundance), due to the strong dilution operating in that case.  

An observational confirmation of these predictions would provide strong
support to the AGB scenario as described by
our models. However, we point out that if observations 
will not find
a large lithium abundance in the blue MS, this would not falsify the AGB
scenario but, rather, it would  simply be an indication that the mass loss rates
adopted for the models of 7.5 and 8\msun are too extreme (see also the
discussion of the models for M~4 in section \ref{m4}).  
These observations are not yet available.

\begin{figure}
\resizebox{.95\hsize}{!}{{\includegraphics{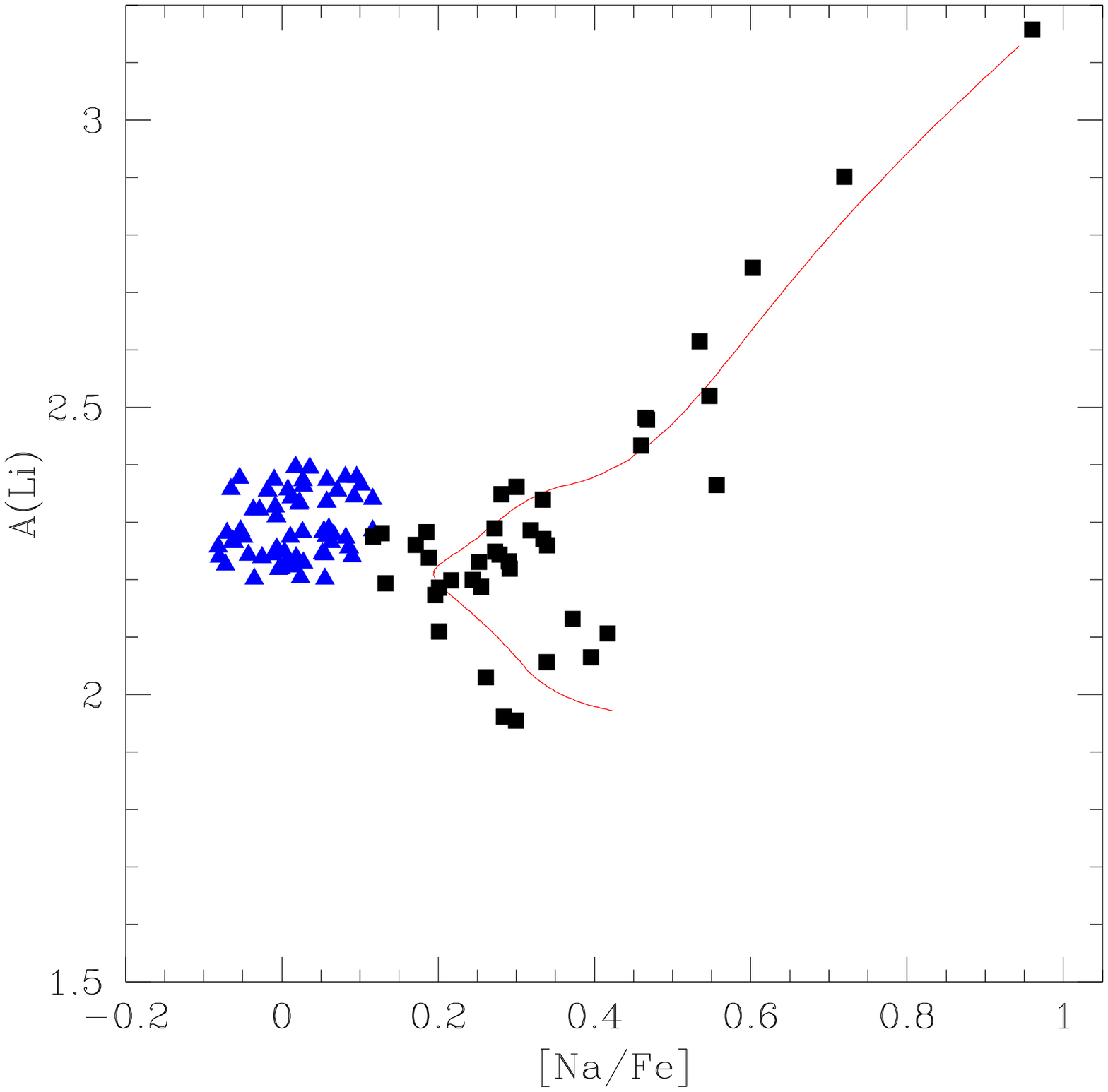} \includegraphics{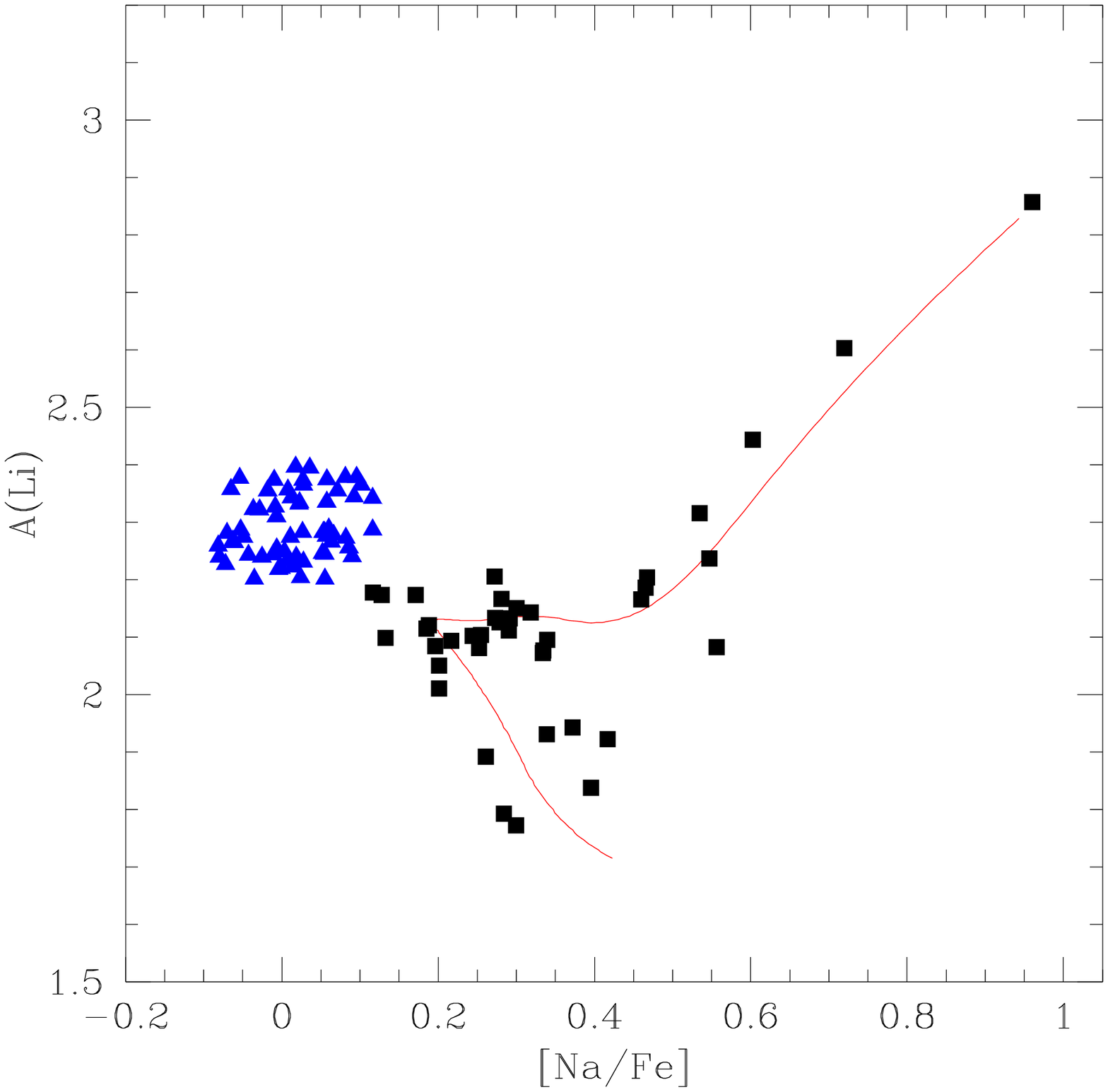}}}
\vskip -40pt
\resizebox{.95\hsize}{!}{{\includegraphics{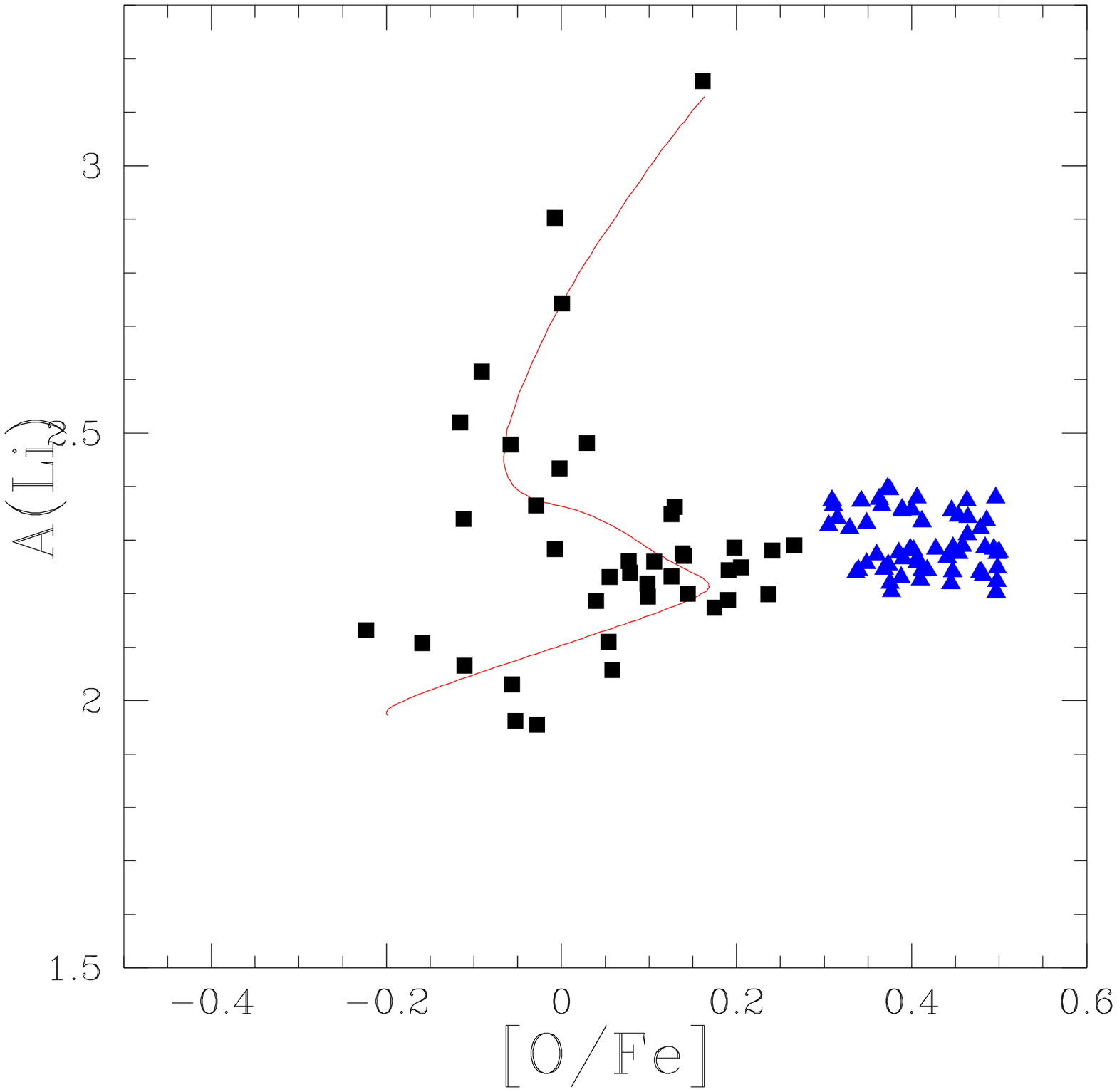} \includegraphics{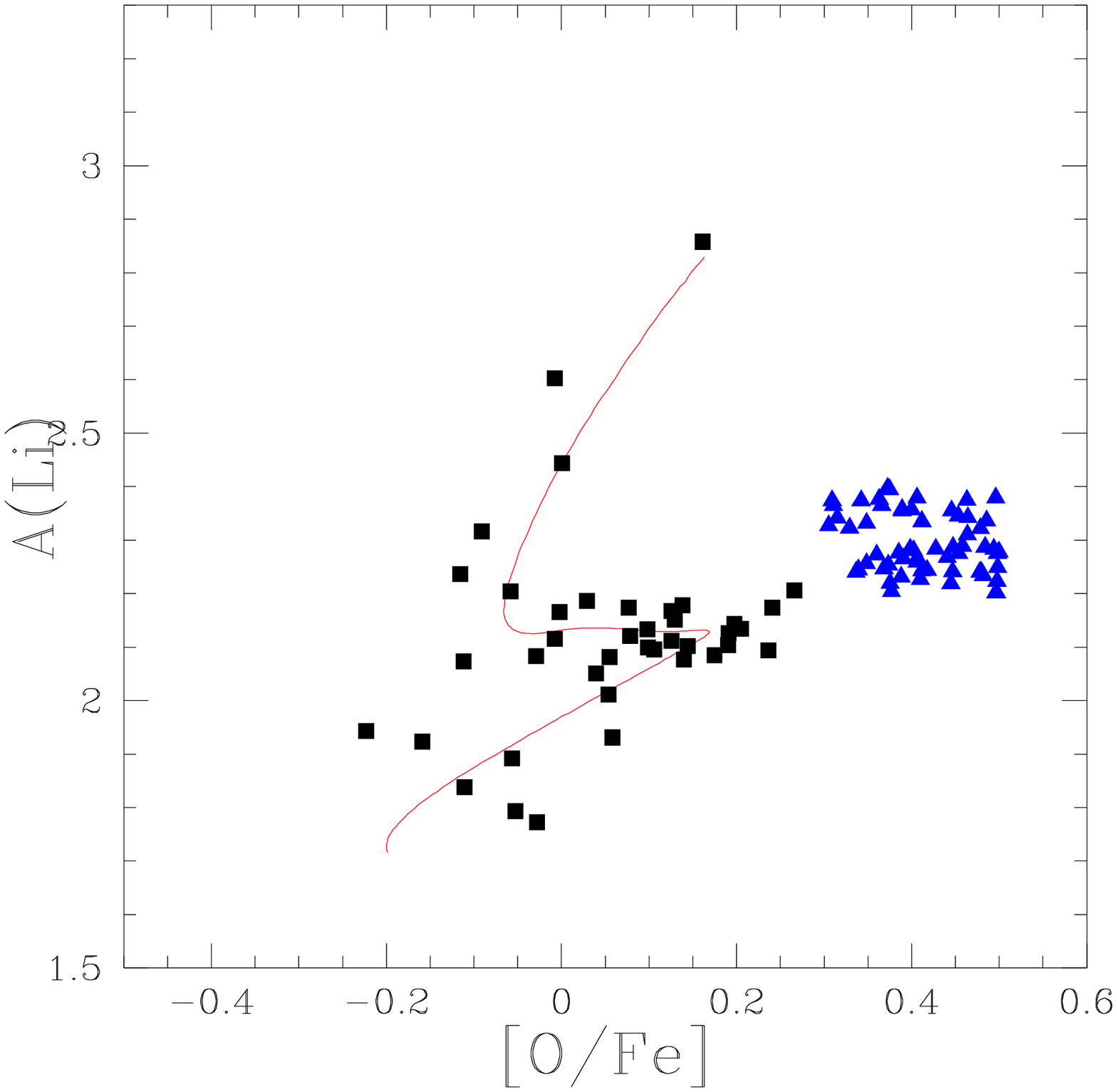}}}
\vskip -25pt
\caption{Li--Na  and Li--O simulations for a cluster like NGC~2808, harboring an extreme, very helium rich, population and an intermediate one. The parameters refer to model NGC~2808-1 in Table 3. We further have A(Li)=2.3 (left panels) and 2.6 (right panels). Blue triangles represent the first generation population, black squares represent the SG. The red line in both panels represents the gas composition along the evolution. }
\label{f3} 
\end{figure}

\begin{figure}
\resizebox{1.\hsize}{!}{\includegraphics{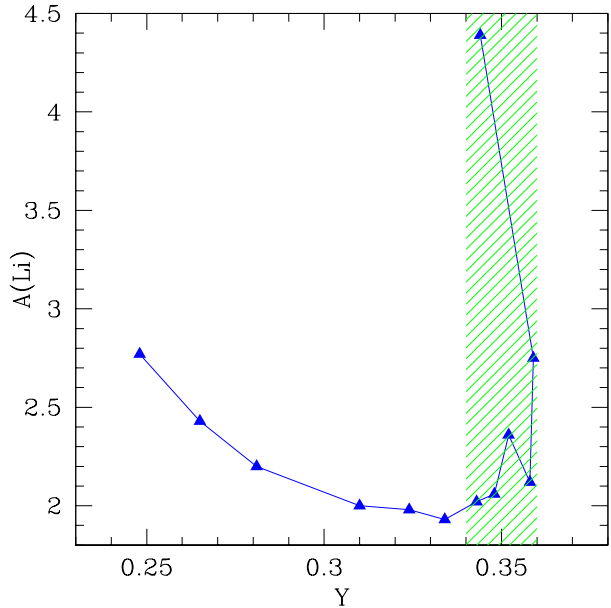}}
\caption{Lithium versus helium content in the ejecta of super--AGB and AGB stars. The dashed area includes masses of 8, 7.5, 7, 6.5, 6, 5.5 from top to bottom: all these ejecta would populate a ``blue MS", if they are not diluted with helium--poor pristine gas before forming SG stars. Only masses from 7.5 to 8\Msun\ provide very lithium rich ejecta.
}
\label{figlihe} 
\end{figure}
\begin{figure}
\resizebox{1.\hsize}{!}{\includegraphics{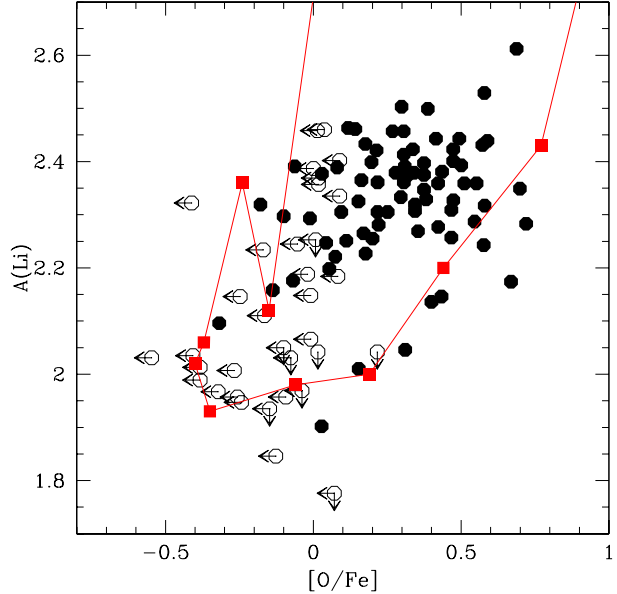}}
\caption{O--Li data for NGC~6752 from Shen et al. 2010. Open symbols represent upper limits. In the plot, we have shifted [O/Fe] by +0.4~dex, as the data in Shen et al. give a standard [O/Fe]$\sim$0 for the FG of this cluster, contrary to the observations by Carretta et al. (2009) that indicate oxygen enhancement for the FG stars in this cluster (see text).  The line connects the (red) squares that show the O--Li yields from Ventura \& D'Antona 2009, 2010 and 2011.} 
\label{f5} 
\end{figure}

It is important to notice, however, that \cite{monaco2010} measured or derived upper limits to the lithium abundances in 91 MS or early subgiant stars in \ocen, finding a remarkable similar value A(Li)=2.19$\pm$0.14 for the whole sample. In their data, a few stars may have A(Li) close to 2.4 -- 2.5 (still within 3$\sigma$\ from the average value), but certainly none has a lithium abundance as large as predicted in Fig.~\ref{f3} for the blue MS stars in NGC~2808. \ocen\ too has a blue MS \citep{bedin2004} that is interpreted to be very helium rich MS \citep[e.g.][]{norris2004}. Observations aimed at selecting specifically blue MS stars to determine their lithium abundance could result in some detections of a very large lithium content. We remark again that high helium does not necessarily mean high lithium, as we show by plotting lithium versus helium content in the ejecta of super--AGB and AGB stars in Fig.~\ref{figlihe}: while masses in the whole dashed area (from 5.5 to 8\msun) may contribute to the blue MS, only stars formed from the ejecta of masses in the range 7.5--8\msun\ would show an outstanding lithium content\footnote{According to \cite{monaco2010}, the absence of stars with high Li abundance in their \ocen\ sample suggests that these were not significantly polluted by lithium produced in AGB stars. This conclusion is not consistent with the models shown here.}

\subsection{The intermediate population} 
Another interesting property of the simulations for NGC~2808 are the abundances of lithium in the intermediate populations: these result from mixing of pristine gas (this latter having a lithium content chosen according to cases A or B) with the lithium of the massive AGBs. 
As shown in Fig.~\ref{f3},  the lithium abundance in the intermediate population is lower than the pristine abundance, and the cases A and B provide interestingly different results. The choice of the pristine gas abundance affects the slope of the average relation Li--Na and Li--O, that are milder in case B.  (see the discussion in Sect.~\ref{6752}). In any case, the finite amount of lithium in the ejecta of massive AGBs produces slopes milder than the slope 1 predicted in the case of pure dilution of pristine gas with lithium free, sodium-rich and oxygen-poor matter.

 \subsection{The case of NGC~6752}
 \label{6752}
 NGC~6752 is the first cluster for which the O--Na anticorrelation has been observed in unevolved stars \citep{gratton2001}, ruling out the possibility that the abundance anomalies are generated by processes occurring inside observed stars, because of the rather low central temperatures and thin convective envelopes of stars at the turn-off of GCs. Afterwards, this finding was confirmed in other clusters, like M 71 \citep[NGC 6838][]{ramirez2002} and 47 Tuc \citep{carretta2004a}.  Further data for giants of NGC~6752 have been collected in \cite{carretta2007a}. The data show a typical O--Na anticorrelation, not extremely extended in oxygen, although several upper limits to [O/Fe] are present. The sodium--poor stars (FG stars) have largely $\alpha$--enhanced  [O/Fe]$\sim$0.4--0.5. 
 
\begin{figure*}
\resizebox{.48\hsize}{!}{\includegraphics{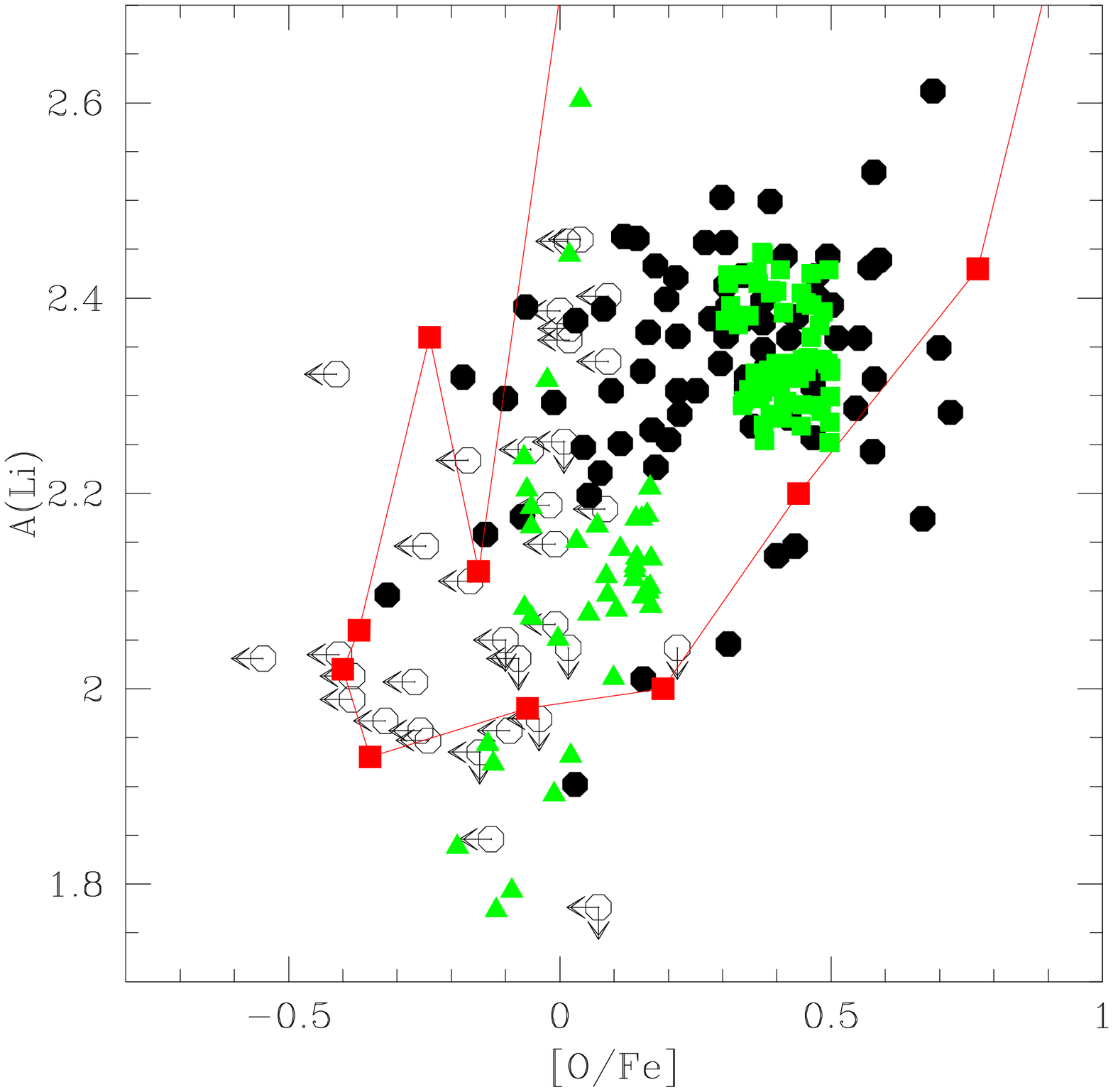}}
\resizebox{.48\hsize}{!}{\includegraphics{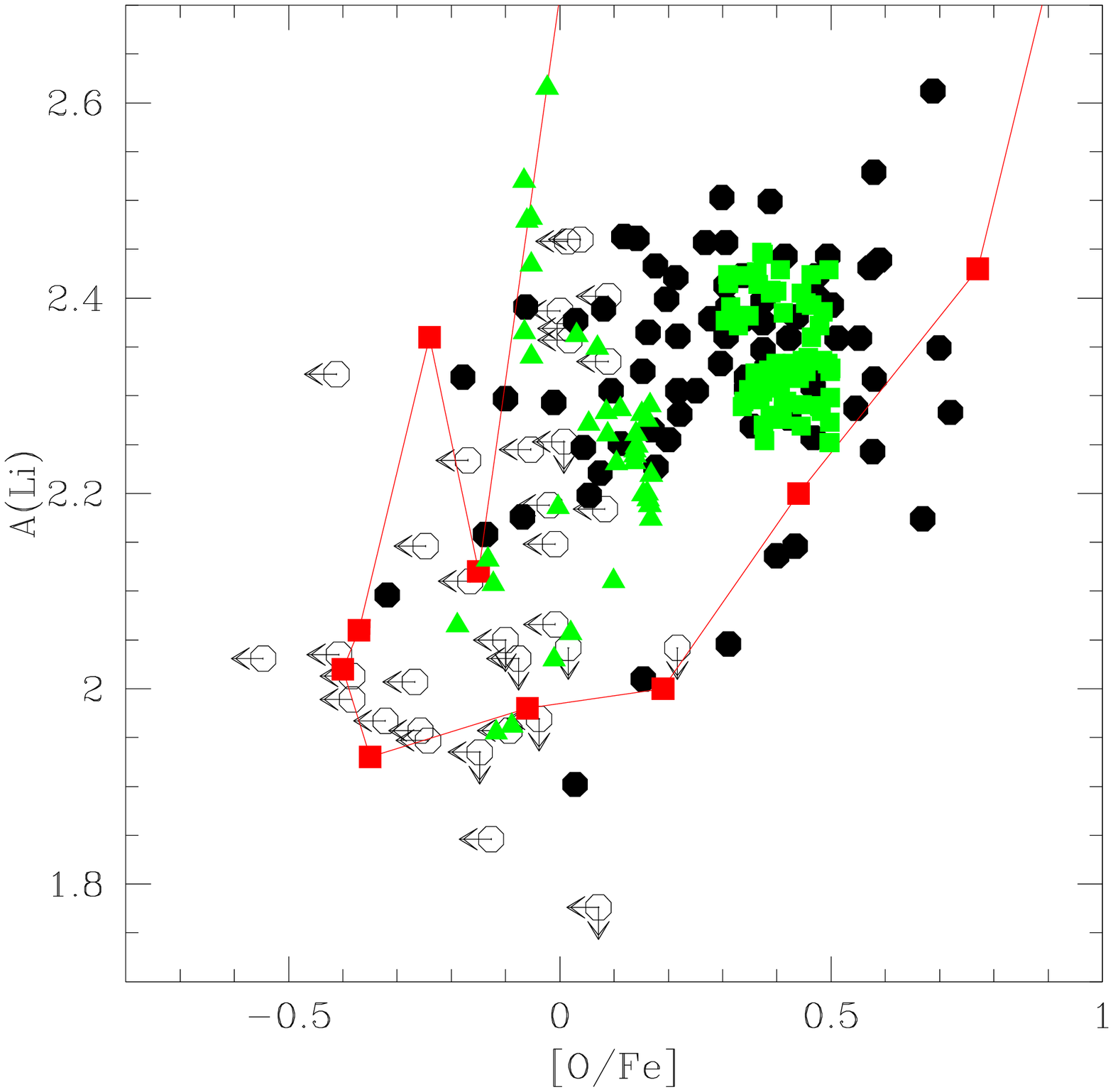}}
%
\caption{Overplotted to the Li--O for NGC~6752 (shifted by +0.4dex, as in Fig. \ref{f5}), we show the results of simulations of Fig.~\ref{f3}. 
Green squares represent the FG and green triangles the SG.
The initial lithium abundance of the diluting pristine gas is A(Li)=2.6 (left panel) and 2.3 (right panel). 
To achieve a better fit of the FG data, we have shifted  A(Li) {\it of the FG} by +0.05~dex. }
\label{fpasq} 
\end{figure*}
  
Lithium in NGC~6752 was measured in nine turnoff stars by \cite{pasquini2005}, who found that its abundance was anticorrelated with sodium and nitrogen, and directly correlated with oxygen. These data formed the basis of an interpretation based on a dilution model of pristine matter with the matter processed in massive rotating stars, having no lithium and high sodium \citep{decressin2007b}. Recently, \cite{shen2010} examined a much larger sample (112 stars), confirming the O--Li correlation, but finding a slope of the correlation $\Delta$A(Li)/$\Delta$[O/Fe]$\simeq$0.4, much milder than the slope 1 expected if the lithium in the pristine matter were just diluted with Na--rich Li--free ejecta. Figure \ref{f5} shows the O--Li data for NGC~6752 together with our yields for super-AGB and AGB stars. Two caveats must be noticed: 1) here we plot the computed abundances, as we deal with turnoff stars, in which no kind of further deep mixing to reduce oxygen can be invoked; 2) in the comparison with simulations, we have shifted the \cite{shen2010} [O/Fe] abundances by +0.4~dex: in fact, the ``standard" [O/Fe] values in \cite{shen2010} analysis is $\sim$0.0--0.1, implying some systematic shift with respect to the abundances of \cite{carretta2007a}.   Our initial models have [O/Fe]=0.4, so the observational values have been adjusted.

The O--Na data for this cluster are not as good as the data for NGC~2808, and also the information of the MS are still incomplete in the literature. Therefore, we use as a guideline the simulations made for NGC~2808.
In Fig.~\ref{fpasq} we overplot the simulations of Fig.~\ref{f3} to the \cite{shen2010} data, after shifting the data by +0.4dex in [O/Fe]. 
We also shift  A(Li) {\it of the FG} by +0.05~dex to achieve a better reproduction of the undepleted abundances in NGC~6752. We adopt the same shift in oxygen for the SG simulation, as the oxygen depletion is directly dependent on the initial abundance, but we do not shift the SG lithium upwards, because the yields in lithium are not linked to the initial abundance chosen for the models, but only to the lithium production through the Cameron Fowler mechanism during the HBB. Fig.~\ref{fpasq}  shows that the abundances in the ejecta may play a role in decreasing the slope of the [O/Fe]--A(Li) anticorrelation.  There is no sign of the few very Li--rich stars predicted for the extreme second generation of NGC~2808, but we see that the ``intermediate" population is broadly consistent with the data. By comparing the two panels we see that  the slope of the simulated data points (excluding the super--AGB values) differs in the two cases (A \& B): as expected, the slope is flatter in the case of A(Li)=2.3. 
We compute an average slope of the simulation by a least square fit both of the FG and SG points, after excluding the extreme lithium abundances. We obtain $\Delta$A(Li)/$\Delta$[O/Fe]=0.27$\pm$0.05 for the case A(Li)=2.3 and $\Delta$A(Li)/$\Delta$[O/Fe]=0.62$\pm$0.05 for the case A(Li)=2.6. 
Thus further observations of the lithium patterns in FG and SG stars providing a more robust determination of the  [O/Fe]--A(Li)  slope, along with a more detailed model in which also other chemical constraints are considered, may even result in a test of the standard BBN.

\section{Models in which the diluting gas does not contain Lithium}
\label{li0}
\begin{figure}
\resizebox{.95\hsize}{!}{{\includegraphics{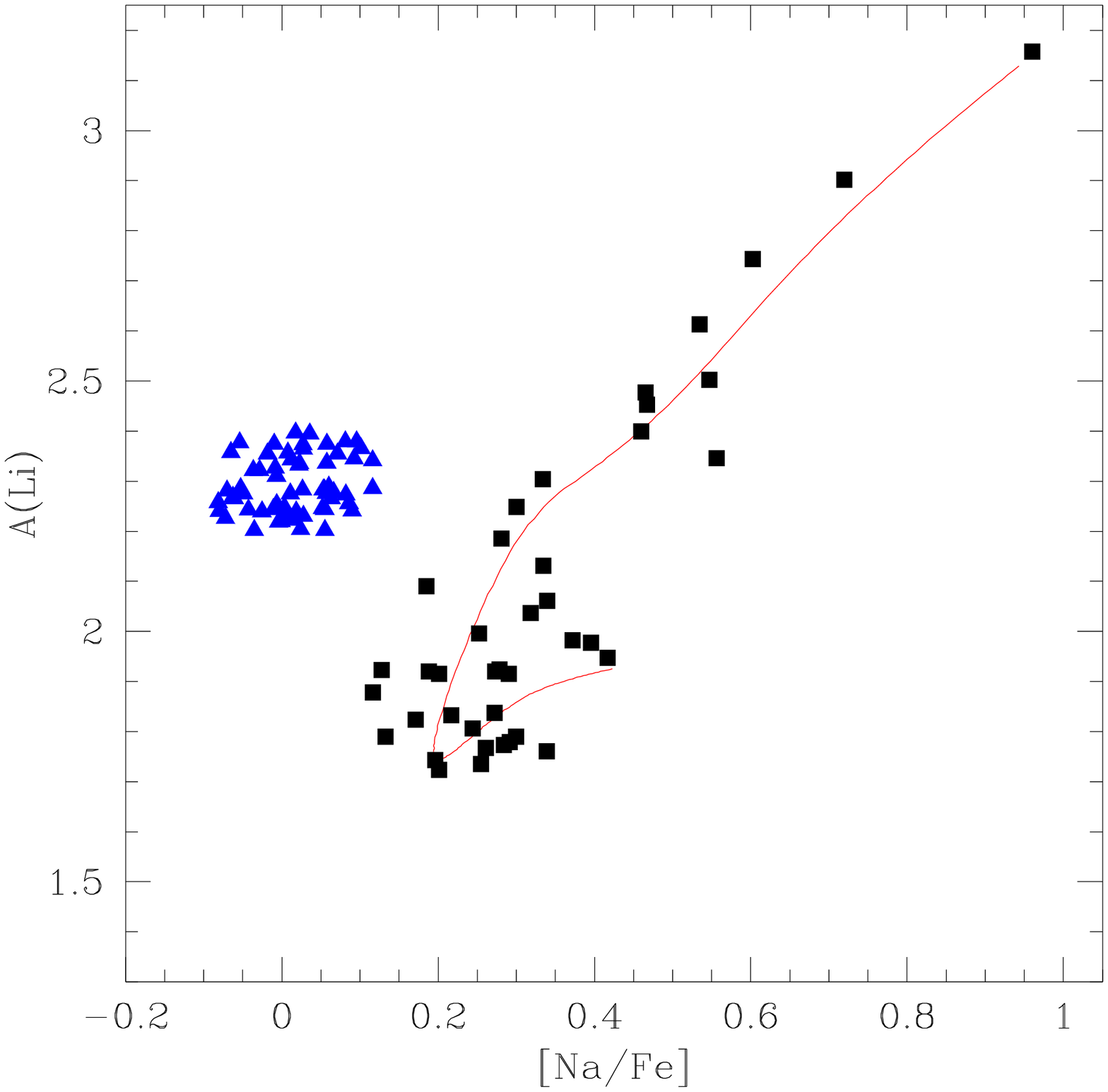} \includegraphics{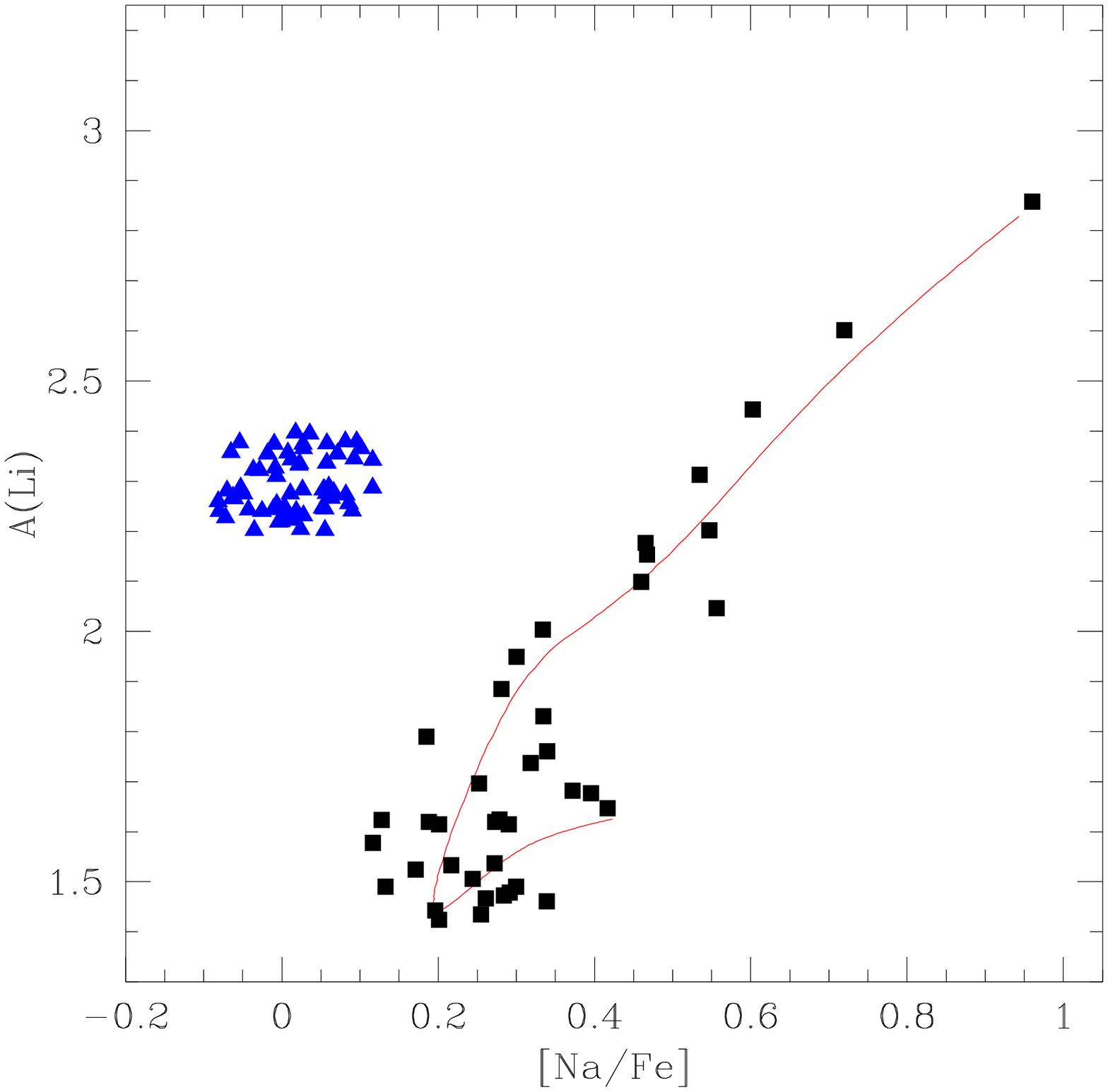}}}
\vskip -40pt
\resizebox{.95\hsize}{!}{{\includegraphics{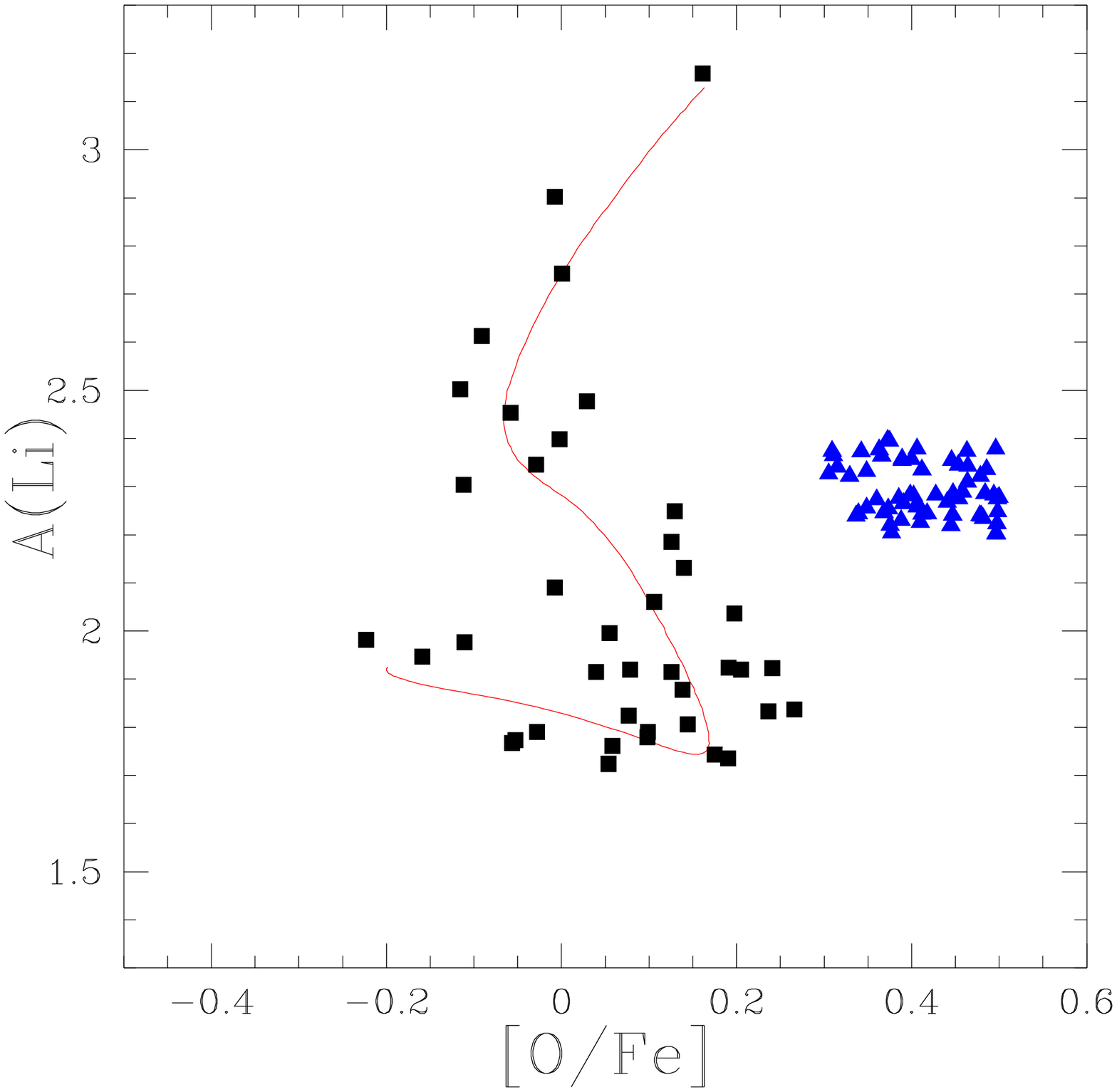} \includegraphics{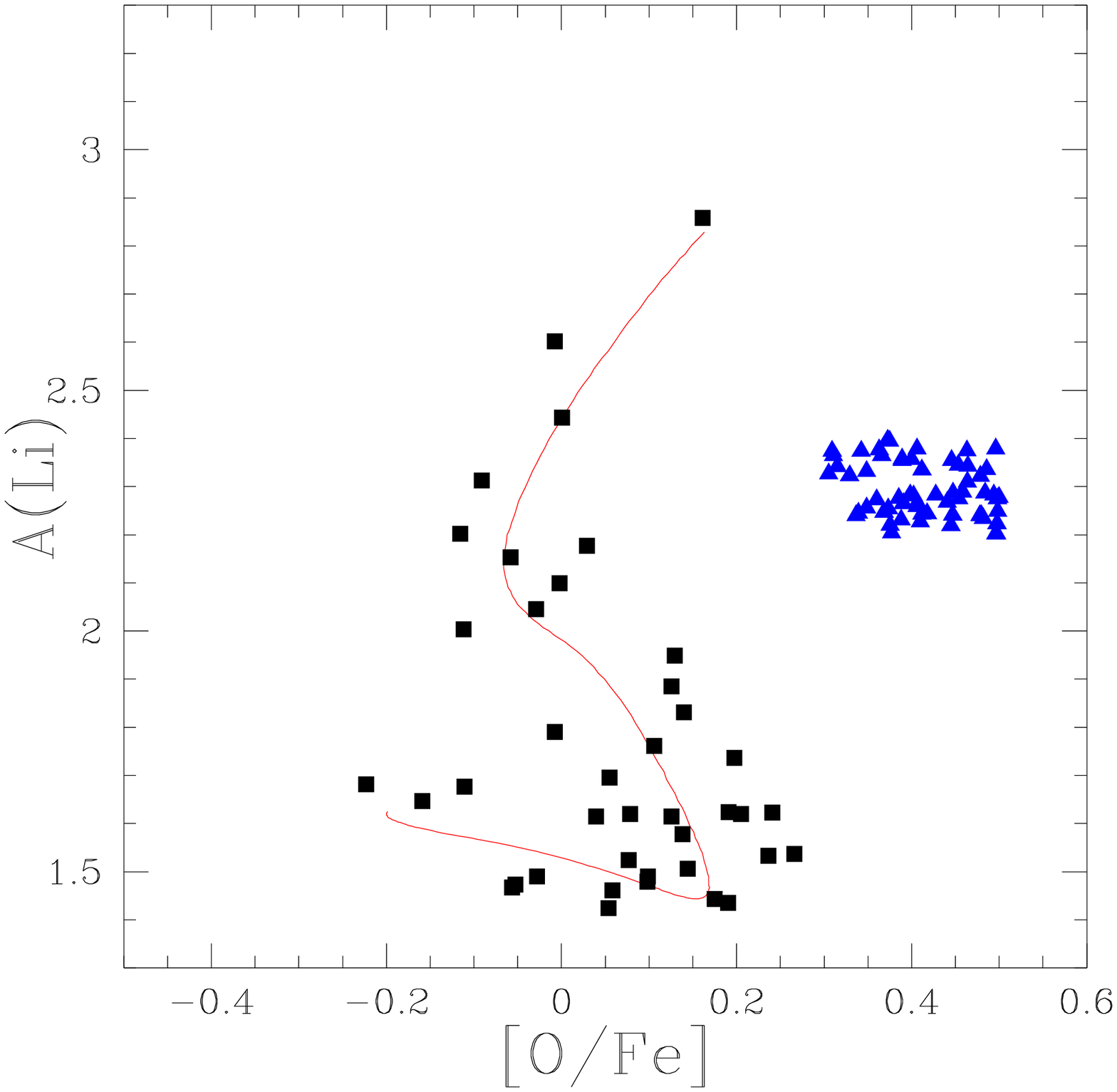}}}
\vskip -25pt
\caption{Li--Na  and Li--O simulations for model NGC~2808-1 in Table 3. The difference with respect to Fig. 5 is in the assumption of no lithium in the {\it diluting} gas. We further have A(Li)=2.3 (left panels) and 2.6 (right panels). Notice that the extreme population is similar to that shown in Fig. 5, because the SG formation occurs in the undiluted matter, while the intermediate population, where dilution is important, has much lower lithium abundances.}
\label{f9} 
\end{figure}

We finally shortly examine case F (Table 3), in which the diluting matter is devoid of lithium. In order to solve the problem of the source of the matter required for dilution, Gratton \& Carretta (2010) proposed that it comes from mass loss from FG stars, that can quantitatively provide the gas necessary to achieve this, if we consider the main sequence stars of all masses and reasonably large mass loss rates. Consequently, this gas has no lithium, because lithium burns below the stellar surface as soon as the temperature increases above $\sim 2 \times 10^6$K, and the mass loss rate required in this scenario implies that layers of matter that have reached this
temperature must contribute to the dilution. Another possible case is when the diluting gas comes from the mass lost by non conservative evolution of massive close binaries \citep{vanbeveren2012} or by close encounters of stars in the dense cores of the GCs (Carini et al. 2012). It is clear that the models for M4 in which there is a strong dilution with pristine matter (from M4-1 to M4-5) lithium in the SG will be very small, if any, in contrast with the observations. As for NGC 2808, in Fig.  \ref{f9} we show the implications of a lithium-free
diluting gas on the Li-O and Li-Na patterns in the model 2808-1. Contrary to the cases shown in Fig. 5, this simulation predicts much smaller lithium abundances for the intermediate population. Further observations of lithium in clusters may help to falsify this model.

\section{The Big Bang lithium abundance}
\label{bbli}
This work, first aimed at further discussing a detailed chemical evolution model for the multiple populations of GCs, has revealed an interesting implication for  the Big Bang nucleosynthesis. In fact,
despite the scarcity of lithium abundance determinations in GCs, these data may result to be a powerful and independent way to constrain the lithium abundance emerging from the Big Bang. The abundance of lithium in the gas forming the SG stars contains information on the abundance of lithium {\it in the gas} in which the hot-CNO processed ejecta are diluted. If we can assume (and this is very reasonable) that, at first order, the possible phenomena that deplete lithium at the surface of population II stars are independent of their being FG or SG stars, the {\it difference} in lithium between FG and SG stars will keep track of the pristine gas abundance. 

In the case of M4, we have seen that the cases B (A(Li)=2.2 in the pristine gas) shown in Fig. 4 can be excluded. This could be a very interesting result, but it is subject to a confirmation of the large lithium yields  shown in Table 2 for super--AGB stars. As far as we have seen from the other observations, there is no present confirmation of those yields. Further, they are theoretically very uncertain, being linked to the choice (unconstrained by observations) of the mass loss rate adopted in these computations. A confirmation could come from observation of large lithium abundances in the blue MS stars of NGC 2808 or \ocen. If in the end the yields of super--AGB stars will be constrained to be much smaller than those used here, case B will still be possible, and we will have to resort to different comparisons.

A very interesting information seems to come from the slope of the O--Li data in NGC 6752: Figure 8 shows the possibility of constraining the primordial lithium abundance from the data. This result is less dependent on the super--AGB yields, as it refers to the (later) phase of formation of the intermediate population. Of course, more and better data, and stronger constraints on the dilution model and on the AGB yields are necessary. We only wish to point out this rewarding possibility, {\it to encourage further and extensive observations of lithium abundances in clusters with multiple populations.}

\section{Discussion}
\label{final}
In this paper we have presented the first predictions for the abundance of Lithium in first and second generation stars of GCs, based on 
quantitative models that can succesfully reproduce the O--Na anticorrelation and the helium abundance distribution in the clusters M~4 and NGC~2808 (Paper~II). Although the model is necessarily parametric, it allows us to discuss the lithium data appeared in the recent literature for several GCs: NGC~6752, M~4, NGC~6397, \ocen. Our  models are based on our recent yield predictions for massive AGB and super--AGB stars.
We show that:
\begin{enumerate}
\item  An important ingredient in the study is the dilution of the ejecta with the pristine gas, and thus whether we interpret the surface abundance of lithium in population II stars as the primordial abundance, or as due to in situ depletion mechanisms. A better understanding of the SG formation could perhaps help to constrain better the Big Bang abundance, although this is at present only wishful thinking.
\item Not all models for M~4 are consistent with the slight decline in lithium by 0.1~dex between the FG and SG. 
A good fit of the data is achieved also assuming zero lithium in the super--AGB matter (rightmost column panels in Fig.\ref{m42}), thanks to the large sodium yields of these ejecta, that require strong dilution with pristine matter. 
Models in which the O--Na ``short" anticorrelation is the result of the rapid formation of SG stars from strongly diluted super--AGB ejecta would solve the mystery of the Be--rich, O--poor star found by \cite{pasquiniber2004} in NGC~6397.
\item In complex clusters, in which a very He--rich extreme population is present, we should expect that this population is born from pure super--AGB ejecta, and thus is {\it very Li--rich}. Observations of lithium in the blue MS of \ocen\ and NGC~2808 may verify or confute this prediction.
\item The O--Li observations of NGC~6752 are consistent with the intermediate population of the models discussed for NGC~2808. The slope of the correlation (milder than the slope 1 expected if the polluting matter is Li--free) is reproduced, thanks to the {\it finite lithium content of the AGB ejecta}. In this case too, a better determination of this slope might provide hints to the Big Bang Lithium abundance.
\item In some models for the SG formation, the diluting gas comes from nuclearly unprocessed  matter from FG stars, that is probably Li--free. We predict that in this case lithium in SG stars should be much smaller than in the FG.
\end{enumerate}

 \section{Acknowledgments} 
This work has been supported through PRIN INAF 2009  "Formation and Early Evolution of Massive Star Cluster" and PRIN INAF 2011  "Multiple populations in Globular Clusters: their role in the Galaxy assembly". EV was supported in part by grant NASA-NNX10AD86G.

\label{lastpage}


\begin{thebibliography}{99}
\bibitem[Asplund et al.(2006)]{asplund2006} Asplund, M., Lambert, D.~L., Nissen, P.~E., Primas, F., \& Smith, V.~V.\ 2006, \apj, 644, 229 

\bibitem[\protect\astroncite{Bedin et al.}{2004}]{bedin2004} Bedin, L.~R., Piotto, G.,  Anderson, J., Cassisi, S., King, I.~R., Momany, Y., \& Carraro, G.\ 2004, ApJ, 605, L125 

\bibitem[Boesgaard et al.(1999)]{boesgaard1999} Boesgaard, A.~M., 
Deliyannis, C.~P., King, J.~R., et al.\ 1999, \aj, 117, 1549 

\bibitem[Carini et al.(2012)]{carini2012} Carini, R. et al.\ 2012, to be submitted 

\bibitem[Carretta et al.(2004)]{carretta2004a} Carretta, E., Gratton, R.~G., Bragaglia, A., Bonifacio, P., \& Pasquini, L.\ 2004, \aap, 416, 925 

\bibitem[Carretta et al.(2007)]{carretta2007a} Carretta, E., Bragaglia, A., Gratton, R.~G., Lucatello, S., \& Momany, Y.\ 2007, \aap, 464, 927 

\bibitem[Carretta et al.(2009)]{carretta2009a} Carretta, E., et al.\ 2009, A\&A, 505, 117 

\bibitem[Cyburt et al.(2008)]{cyburt2008} Cyburt, R.~H., Fields, B.~D., \& Olive, K.~A.\ 2008, J. Cosmol. Astro-Part. Phys., 11, 12 

\bibitem[\protect\citeauthoryear{D'Antona et al.}{2002}]{dantona2002} D'Antona, F., Caloi, V., Montalb\'{a}n, J., Ventura, P., \& Gratton, R.~2002, A\&A, 395, 69

\bibitem[D'Antona \& Caloi(2004)]{dc2004} D'Antona, F., \& Caloi, V.\ 2004, \apj, 611, 871 

\bibitem[Decressin et al.(2007a)]{decressin2007a} Decressin, T.,  Meynet, G., Charbonnel, C., Prantzos, N., \& Ekstr{\"o}m, S.\ 2007a, A\&A, 464, 1029 

\bibitem[Decressin et al.(2007b)]{decressin2007b} Decressin, T., Charbonnel, C., \& Meynet, G.\ 2007, \aap, 475, 859 

\bibitem[de Mink et al.(2009)]{demink2009} de Mink, S.~E., Pols, O.~R., Langer, N., \& Izzard, R.~G.\ 2009, \aap, 507, L1 

\bibitem[D'Ercole et al.(2008)]{dercole2008} D'Ercole, A., 
Vesperini, E., D'Antona, F., McMillan, S.~L.~W., \& Recchi, S.\ 2008, \mnras, 391, 825 

\bibitem[D'Ercole et al.(2010)]{dercole2010} D'Ercole, A., D'Antona, F., Ventura, P., Vesperini, E., 
\& McMillan, S.~L.~W.\ 2010, \mnras, 407, 854 

\bibitem[D'Ercole et al.(2011)]{dercole2011} D'Ercole, A., D'Antona, F., \& Vesperini, E.\ 2011, \mnras, 415, 1304 

\bibitem[D'Ercole et al.(2012)]{dercole2012cev1} D'Ercole, A., et al.\ 2012, \mnras, 423, 1521 (Paper II)


\bibitem[D'Orazi et al.(2010)]{dorazi47tuc2010} D'Orazi, V., Lucatello, S., Gratton, R., Bragaglia, A., Carretta, E., Shen, Z., \& Zaggia, S.\ 2010, \apjl, 713, L1 

\bibitem[D'Orazi \& Marino(2010)]{dorazi2010} D'Orazi, V., \& Marino, A.~F.\ 2010, \apjl, 716, L166 

\bibitem[Fields(2011)]{fields2011} Fields, B.~D.\ 2011, Annual Review of Nuclear and Particle Science, 61, 47 

\bibitem[\protect\citeauthoryear{Gratton et al.}{2001}]{gratton2001}Gratton, R.~G., Bonifacio, P., Bragaglia, A., et al.~2001, A\&A, 369, 87

\bibitem[Gratton \& Carretta(2010)]{gratton2010} Gratton, R.~G., \& Carretta, E.\ 2010, \aap, 521, A54 

\bibitem[Iocco et al.(2009)]{iocco2009} Iocco, F., Mangano, G., Miele, G., Pisanti, O., \& Serpico, P.~D.\ 2009, Phys. Rep., 472, 1 

\bibitem[Karakas et al.(2006)]{karakas2006} Karakas, A.~I., Fenner, Y., Sills, A., Campbell, S.~W., \& Lattanzio, J.~C.\ 2006, \apj, 652, 1240 

\bibitem[Koch et al.(2011)]{koch2011} Koch, A., Lind, K., \& Rich, R.~M.\ 2011, \apjl, 738, L29 

\bibitem[Lind et al.(2009)]{lind2009} Lind, K., Primas, F., Charbonnel, C., Grundahl, F., \& Asplund, M.\ 2009, \aap, 503, 545 

\bibitem[Lind et al.(2011)]{lind2011} Lind, K., Charbonnel, C., Decressin, T., Primas, F., Grundahl, F., \& Asplund, M.\ 2011, \aap, 527, A148 

\bibitem[Mel{\'e}ndez \& Ram{\'{\i}}rez(2004)]{melendez2004} Mel{\'e}ndez, J., \& Ram{\'{\i}}rez, I.\ 2004, \apjl, 615, L33 

\bibitem[Meynet et al.(2006)]{meynet2006} Meynet, G., Ekstr{\"o}m, S., \& Maeder, A.\ 2006, A\&A, 447, 623 

\bibitem[Milone et al.(2010)]{milone2010} Milone, A.~P., Piotto, G., King, I.~R., et al.\ 2010, \apj, 709, 1183 


\bibitem[Monaco et al.(2010)]{monaco2010} Monaco, L., Bonifacio, P., Sbordone, L., Villanova, S., \& Pancino, E.\ 2010, \aap, 519, L3 

\bibitem[Monaco et al.(2012)]{monaco2012} Monaco, L., Villanova, S., Bonifacio, P., et al.\ 2012,  \aap, 539, A157 (arXiv:1108.0138) 

\bibitem[Monaco et 
al.(2012)]{2012A&A...539A.157M} Monaco, L., Villanova, S., Bonifacio, P., et al.\ 2012, \aap, 539, A157 


\bibitem[\protect\astroncite{Norris}{2004}]{norris2004} Norris, J.~E.\ 2004, ApJ, 612, L25 

\bibitem[Pasquini et al.(2004)]{pasquiniber2004} Pasquini, L., Bonifacio, P., Randich, S., Galli, D., \& Gratton, R.~G.\ 2004, \aap, 426, 651 

\bibitem[Pasquini et al.(2005)]{pasquini2005} Pasquini, L., Bonifacio, P., Molaro, P., Francois, P., Spite, F., Gratton, R.~G., Carretta, E., \& Wolff, B.\ 2005, \aap, 441, 549 

\bibitem[Piau et al.(2006)]{piau2006} Piau, L., Beers, T.~C., Balsara, D.~S., Sivarani, T., Truran, J.~W., \& Ferguson, J.~W.\ 2006, \apj, 653, 300 

\bibitem[Prantzos \& Charbonnel(2006)]{prantzos2006} Prantzos, N., \& Charbonnel, C.\ 2006, \aap, 458, 135 

\bibitem[Ram{\'{\i}}rez \& Cohen(2002)]{ramirez2002} Ram{\'{\i}}rez, S.~V., \& Cohen, J.~G.\ 2002, \aj, 123, 3277 

\bibitem[Renzini(2008)]{renzini2008} Renzini, A.\ 2008, \mnras, 391, 354 

\bibitem[Sbordone et al.(2010)]{sbordone2010} Sbordone, L., et al.\ 2010, \aap, 522, A26 

\bibitem[Shen et al.(2010)]{shen2010} Shen, Z.-X., Bonifacio, P., Pasquini, L., \& Zaggia, S.\ 2010, \aap, 524, L2 

\bibitem[Sills \& Glebbeek(2010)]{sills2010} Sills, A., \& Glebbeek, E.\ 2010, \mnras, 407, 277 

\bibitem[Spergel et al.(2007)]{spergel2007} Spergel, D.~N., et al.\ 2007, \apjs, 170, 377 

\bibitem[Vanbeveren et al.(2012)]{vanbeveren2012} Vanbeveren, D., Mennekens, N., \& De Greve, J.~P.\ 2012, A\&A, in press  (arXiv:1109.2713) 

\bibitem[\protect\citeauthoryear{Ventura et al.}{2001}]{ventura2001} Ventura, P., D'Antona, F., Mazzitelli, I., \& Gratton, R.\ 2001, ApJ, 550, L65 

\bibitem[\protect\citeauthoryear{Ventura et al.}{2002}]{vdm2002} Ventura, P., D'Antona, F., \& Mazzitelli, I.\ 2002, A\&A, 393, 215 

\bibitem[Ventura \& D'Antona(2009)]{vd2009} Ventura P., D'Antona F., 2009, A\&A, 499, 835

\bibitem[Ventura \& D'Antona(2010)]{vd2010litio} Ventura, P., \& D'Antona, F.\ 2010, \mnras, 402, L72 

\bibitem[Ventura \& D'Antona(2011)]{vd2011} Ventura, P., \& D'Antona, F.\ 2011, \mnras, 410, 2760 


\end{thebibliography}
\end{document}